\documentclass[12pt]{article}

\usepackage{amssymb,epsfig}

\begin{document}

\begin{center}
\baselineskip=24pt

{\Large \bf Neutron background in large-scale xenon detectors
for dark matter searches}
\vspace{0.5cm}

{\large
M. J. Carson, J. C. Davies, E. Daw, R. J. Hollingworth,
V. A. Kudryavtsev~\footnote{Corresponding author, 
e-mail: v.kudryavtsev@sheffield.ac.uk}, T. B. Lawson,
P. K. Lightfoot, J. E. McMillan, B. Morgan, 
S. M. Paling, M. Robinson, 
N. J. C. Spooner~\footnote{Corresponding author, 
e-mail: n.spooner@sheffield.ac.uk}, 
D. R. Tovey}

\vspace{0.5cm}

{\it Department of Physics and Astronomy, 
University of Sheffield, Sheffield S3 7RH, UK}

\vspace{0.5cm}
\begin{abstract}
Simulations of the neutron background for
future large-scale particle dark matter detectors are presented.
Neutrons were generated in rock and detector elements 
via spontaneous fission and ($\alpha$,n) reactions, and by 
cosmic-ray muons. The simulation techniques and results are discussed
in the context of the expected sensitivity of a generic liquid xenon dark matter 
detector. Methods of neutron background suppression are investigated.
A sensitivity of $10^{-9}-10^{-10}$ pb to WIMP-nucleon interactions
can be achieved by a tonne-scale detector.
\end{abstract}

\end{center}

\vspace{0.5cm}


\vspace{0.5cm}
\noindent Corresponding authors: V. A. Kudryavtsev, N. J. C. Spooner,
Department of Physics and Astronomy, University of Sheffield, 
Hicks Building, Hounsfield Road, 
Sheffield S3 7RH, UK

\noindent Tel: +44 (0)114 2224531; \hspace{2cm} Fax: +44 (0)114 2728079; 

\noindent E-mail: v.kudryavtsev@sheffield.ac.uk, n.spooner@sheffield.ac.uk

\pagebreak

{\large \bf 1. Introduction}
\vspace{0.3cm}

\begin{figure}[htb]
\begin{center}
\epsfig{figure=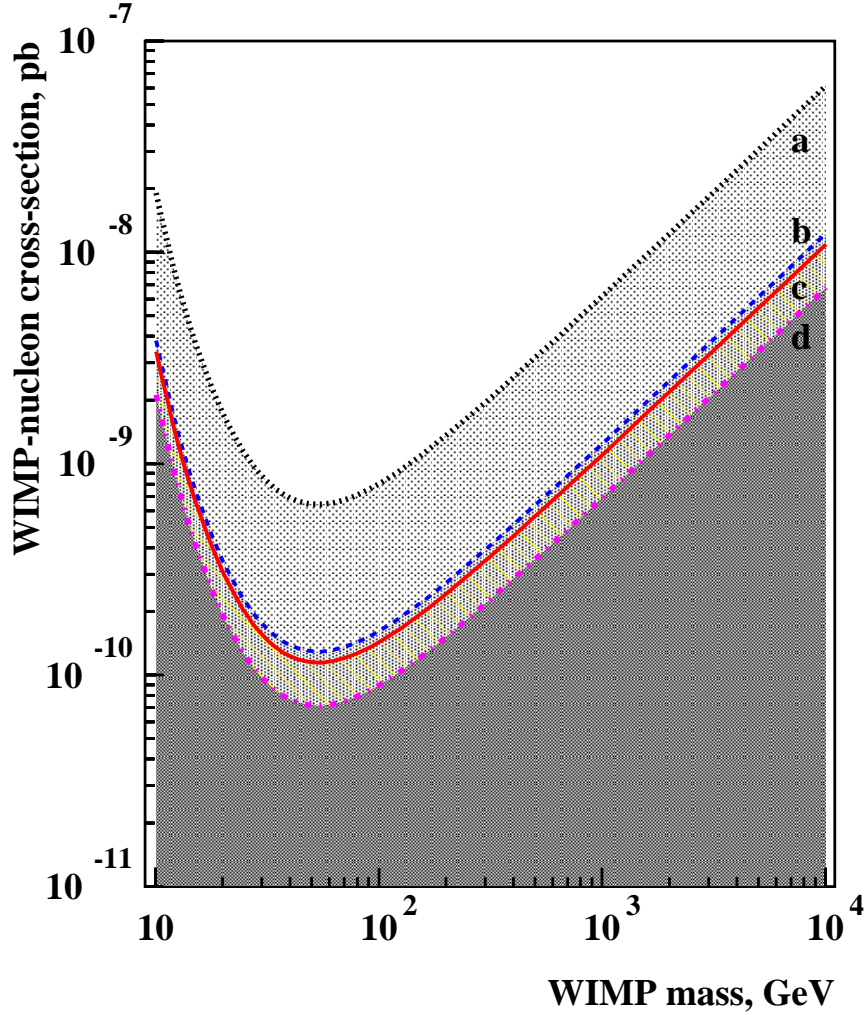,height=15cm}
\caption{Limitations on the sensitivity of a future
250 kg xenon detector (one year running time) for dark matter searches
assuming 100\% rejection of gamma and alpha induced events. For any
specific detector configuration (see text for details) the parameter
space below the curve cannot be probed because of the neutron 
background.
Dotted curve (a) -- detector with 169 2-inch 
ultra-low-background R8778
PMTs surrounded by hydrocarbon passive shielding (Configuration 1). 
Dashed curve (b) -- detector with liquid
scintillator veto system (Configuration 3); background is statistically subtracted. 
Solid curve (c) --
detector with an active veto assuming further improvements
in a PMT design (large PMTs with lower contamination levels, Configuration 
5). Dashed-dotted curve (d) -- ultimate limit
for a detector with no background events observed during
one year of running reachable with charge readout
and ultra-pure materials.}
\label{fig-sens}
\end{center}
\end{figure}

\indent Future dark matter experiments planning to reach a sensitivity
of $10^{-9}-10^{-10}$ pb to the WIMP-nucleon cross-section
require a very low background environment,
sophisticated techniques capable of 
discriminating WIMP-induced events from all kinds of background
and at least one tonne of target mass to achieve sufficient counting rate. 
Hereafter, the quoted sensitivity refers to the minimum
of the sensitivity curve, which occurs for WIMP masses in the range 40-80 GeV
(see, for instance, Figure \ref{fig-sens} for examples of the
sensitivity or exclusion curves). 
Some individual background events, 
however, are indistinguishable from expected WIMP scattering events.  
WIMPs are expected to interact with ordinary 
matter in detectors to produce nuclear recoils, which can be 
detected through ionisation, scintillation or phonons. Identical 
events can be induced by single elastic scattering of neutrons. 
Thus, only suppression of any background
neutron flux by passive or active shielding and proper choice of
detector materials will allow experiments to 
reach sufficiently high sensitivity to WIMPs. Designing detectors, 
their shielding and active veto systems 
requires simulation of neutron fluxes from various sources.

Neutrons underground arise from two sources: i) local 
radioactivity, and ii) cosmic-ray muons. Neutrons associated with 
local radioactivity are produced mainly via ($\alpha$,n) reactions 
initiated by $\alpha$-particles from U/Th traces in the rock and 
detector elements. Neutrons from spontaneous fission of $^{238}$U 
also contribute to the flux. The neutron yield from 
cosmic-ray muons depends strongly on the depth of the underground 
laboratory. The suppression of the muon flux by a large 
rock overburden will also reduce the neutron flux but by a smaller
factor.

In the present study neutrons associated with radioactivity in
rock and detector elements are treated separately. Although these
neutrons come from similar reactions, the materials in which they are produced
are certainly different, as are the methods of their suppression.
Neutrons from surrounding rock can be easily suppressed by passive 
hydrocarbon shielding, whereas the internal neutron flux can be 
reduced by choosing ultra-low-background
materials and possibly by using an active veto to reject
events in the detector in coincidence with veto signals.

At deep underground sites (3 km w.e. or more), the 
neutron production rate from muons is about 3 orders of magnitude lower
than the rate for neutrons arising from rock activity, depending strongly 
both on the depth and the U/Th contamination. 
The muon-induced neutron flux can
be important, however, for experiments intending to reach high 
sensitivity to WIMPs. There are 
several reasons for this: 1) the energy spectrum of muon-induced neutrons 
is hard, extending to GeV energies, and fast neutrons can travel far from 
the associated
muon track, reaching a detector from large distances; 2) fast neutrons 
transfer larger energies to nuclear recoils making them visible in 
dark matter detectors, while many recoils from $\alpha$-induced neutrons
fall below detector energy thresholds; 3) a detector 
can be protected against neutrons from the rock activity by 
hydrocarbon material, possibly with the addition of a thermal neutron 
absorber; such material, however, will also be a target for cosmic-ray 
muons. 

This work includes, for the first time,
a detailed Monte Carlo simulation of production, propagation and detection
of neutrons from known sources, investigation of techniques for 
neutron flux suppression and studies of systematics associated
with the neutron background in connection
with the sensitivity of a future detector to WIMP-nucleus
interactions. The work is part of a programme of
neutron background studies for the dark matter experiments at Boulby 
mine (North Yorkshire, UK) (see Ref. \cite{boulby} for a review of
dark matter searches at Boulby). Similar studies have
been initiated for dark matter
experiments at Gran Sasso and Modane \cite{cresst,chardin}.
The present simulations were carried out
for a large-scale xenon detector and are relevant to several
programmes around the world, including other potential large-scale
dark matter detectors.
A tonne-scale xenon dark matter detector is planned for the
Boulby Underground Laboratory \cite{zeplinmax}. 
A similar detector has been proposed for a new
underground laboratory in the USA \cite{xenon}. Another
liquid xenon based detector is XMASS II \cite{xmass} to be built in Japan for
solar neutrino, double-beta decay and dark matter searches.
The double-beta decay experiment EXO \cite{exo} will also be based on 
liquid xenon. The simulations presented here are important for many
detectors designed for rare event studies.

The paper is organised as follows. 
Generation of neutron spectra from ($\alpha$,n) reactions is
described in Section 2.
Neutrons from rock and the
required shielding are discussed in Section 3. 
Simulations of muon-induced neutrons are presented in Section 4. 
Neutron background from detector elements is investigated in Section 5.
Systematic effects caused by neutron background in connection with the
sensitivity of a large-scale xenon detector to WIMP-nucleon cross-section
are studied in Section 6.
The summary and conclusions are given in Section 7.

\vspace{0.5cm}
{\large \bf 2. Neutron production by radioactive isotopes}
\vspace{0.3cm}

Neutron production by radioactive isotopes in the decay chains of uranium and
thorium was calculated using the SOURCES code \cite{sources}.
The main features of the code are as follows. 
Spontaneous
fission (of $^{238}$U mainly) was simulated using a Watt spectrum 
\cite{watt}.
Neutron fluxes and spectra from ($\alpha$,n) reactions were obtained
taking into account the lifetimes of isotopes, energy spectra of alphas,
cross-sections of reactions as functions of alpha energy,
branching ratios for transitions to different excited states,
stopping power of alphas in various media,
and assuming isotropic emission of neutrons in the centre-of-mass system.

SOURCES provides a treatment of ($\alpha$,n) reactions only up to 6.5 MeV 
$\alpha$-energies. This is likely to restrict significantly the reliability of
the results because the cross-sections of ($\alpha$,n) reactions rise with
energy and the average neutron energy also increases 
with the parent alpha energy. Hence, the 6.5 MeV cut reduces the total
neutron yield from ($\alpha$,n) reactions and artificially shifts the neutron 
spectrum to lower energies. The effect of the neutron spectrum shift can be
significant. We tested this by generating neutron spectra in NaCl with the
original SOURCES code and by taking a different spectrum from Ref. \cite{modane},
calculated for the Modane rock.
We then propagated neutrons with both of these spectra
through lead and various thicknesses of
hydrocarbon material (CH$_2$) usually used to shield detectors from rock neutrons
(see Section 3 for details of propagation procedure), and
compared the two results. Even if both spectra
are normalised to the same neutron production rate, there remains 
about a two order of magnitude difference
in the predicted neutron
flux above 10-100 keV after 30 cm of lead and 35 g/cm$^2$ of hydrocarbon,
the spectrum from SOURCES giving a lower rate because of the smaller neutron
energies. The effect is mainly due to the decrease in neutron-proton elastic
scattering cross-section with energy. It became obvious that
the neutron production code had to be modified to provide a more realistic
treatment of ($\alpha$,n) reactions. Note that the neutron production 
was simulated in Ref. \cite{modane} assuming a transition of the nucleus
to the ground state only (this overestimates the neutron energy) and
an emission of a neutron at 90$^{\circ}$ (which means that neutron energy
was directly calculated from the alpha energy and the neutron spectrum was a
delta-function).

The following modifications were made to SOURCES to overcome the
6.5 MeV limit. 
Existing cross-sections were extended to 10 MeV, taking into account
available experimental data. For some materials, new cross-sections
were added to the code. For example, we added the cross-section for
$^{23}$Na measured up to 10 MeV \cite{na} as an alternative to those
already present in the code library. The cross-section for $^{35}$Cl was
not present initially in the code library and was added from Ref. \cite{cl}.
The cross-sections on Na and Cl were needed for calculation
of neutron production in the salt rock. If the cross-section
for a material was measured or calculated for low energies only, then
it was extrapolated from low energies up to 10 MeV. The Nuclear Data
Services of the Nuclear Data Centre at the International Atomic Energy
Agency \cite{iaea} were extensively used to obtain cross-sections.

The branching ratios for transitions to the ground and excited states
above 6.5 MeV were chosen to be the same as at 6.5 MeV. This resulted in a 
small overestimate of neutron energies for alphas above 6.5 MeV,
since the increased probability of transition to the higher states
was neglected, but the total neutron flux was not affected. The 
uncertainties associated with the calculations of such probabilities
were not negligible, however. The differences as large as (20-30)\% exist 
between different calculations of the transition probabilities in the 
SOURCES library. If the excited levels were not in
the code library, as was the case for elements for which 
the cross-sections were absent too, then in adding the cross-section
we assumed that the transition was occuring to the ground state only.

\vspace{0.5cm}
{\large \bf 3. Neutrons from rock}
\vspace{0.3cm}

Here, our main objective was to find the thickness of 
hydrocarbon shielding needed to suppress the neutron flux from rock 
down to a level allowing the required sensitivity to 
WIMP-nucleus interactions. We started with neutron production in rock,
then we propagated neutrons through the rock to the rock/cavern boundary and
further on through lead and hydrocarbon shielding to the detector.
Finally we generated nuclear recoils from neutrons in the xenon target 
within the detector.

\begin{figure}[htb]
\begin{center}
\epsfig{figure=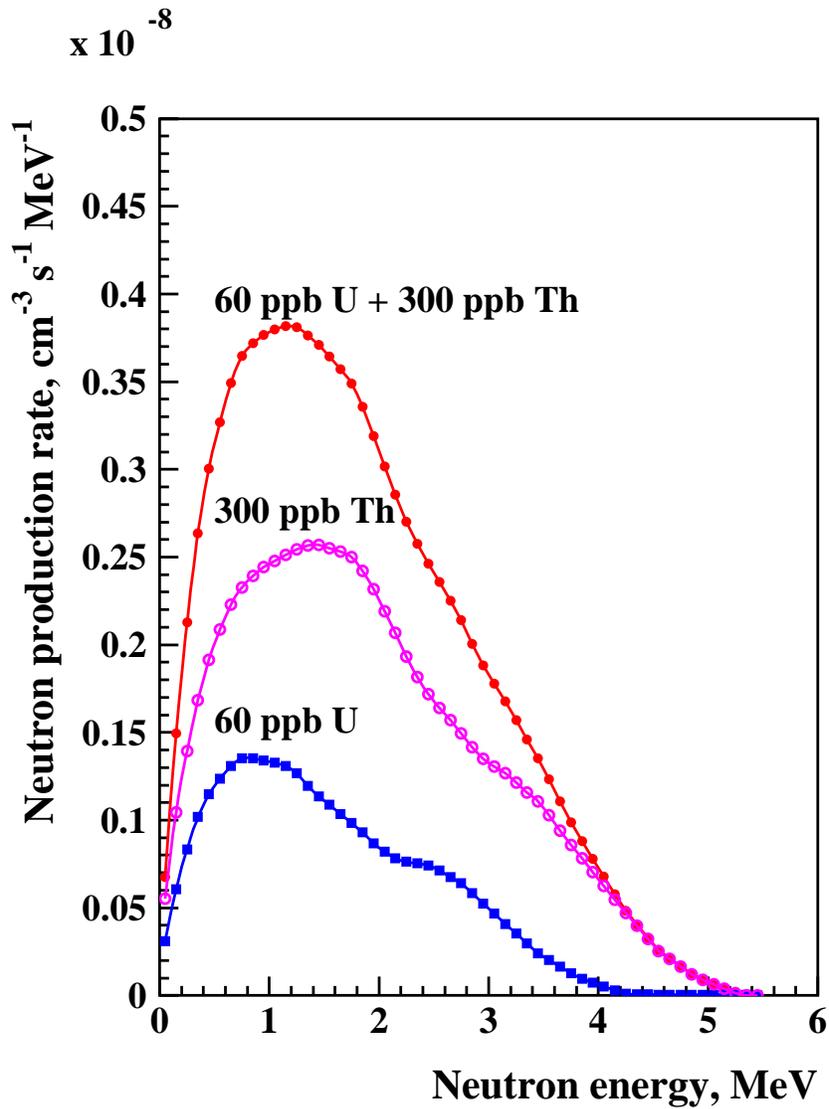,height=15cm}
\caption{Neutron energy spectrum from U and Th traces in rock 
as calculated with modified SOURCES. Contributions from 60 ppb U
(filled squares and lower curve), 300 ppb Th (open circles and 
middle curve) and the sum of the two (filled circles and upper
curve) are shown.} 
\label{fig-nprod}
\end{center}
\end{figure}

Simulation of neutron production 
via spontaneous fission and ($\alpha$,n) reactions in rock was carried out 
with the modified SOURCES code (see Section 2). Rock was assumed to be halite 
(NaCl), which is the case for the Boulby Underground Laboratory (UK) 
and the Waste Isolation Pilot Plant at Carlsbad (USA), both being the 
proposed sites for underground experiments. The 
contamination levels of radioactive elements in rock vary from site to site
and can vary also from hall to hall within an underground laboratory.
In these simulations they were taken as 60 ppb of U 
and 300 ppb of Th in secular equilibrium.
The energy spectrum of neutrons at production
from SOURCES is shown in Figure \ref{fig-nprod}.
The total neutron production rate was found to be about $1.05 \times 10^{-7}$
cm$^{-3}$ s$^{-1}$.
The neutron energy spectra in NaCl obtained with SOURCES 
are similar in shape for U and Th initiated neutrons 
(see Figure \ref{fig-nprod}). 
This means that for other contamination levels of U and Th the spectrum of 
neutrons (as well as the spectrum of nuclear recoils in a detector) 
can be scaled from that reported here taking into account the difference 
in contamination levels. For equal concentrations in NaCl, uranium gives roughly 
twice as many neutrons as thorium. For 100 ppb of uranium the neutron
production rate is equal to $5.2 \times 10^{-8}$
cm$^{-3}$ s$^{-1}$, whereas for a 100 ppb of thorium 
it is $2.5 \times 10^{-8}$
cm$^{-3}$ s$^{-1}$.
Decreasing the Th level down to 150 ppb (50\% of the basic value 
used in the present work) results in a decrease of the neutron yield down 
to about 65\% of the initial value with only a tiny decrease of the mean
neutron energy from 1.81 MeV down to 1.75 MeV.

Neutron propagation through the rock was simulated using the GEANT4 
package \cite{geant4}.
Neutrons from the rock wall were produced in a slab of rock 
$1 \times 1$ m$^{2}$ with
3 m depth into the rock (simulations with varying rock thicknesses
showed that only those neutrons within 3 m of the rock surface are 
capable of reaching it). Neutrons from this region were allowed to
propagate isotropically into a much larger region ($100 \times 100$ m$^{2}$
also with 3 m depth). This avoids neutron losses
due to rock edge effects. The total spectrum observed from this region was
then re-scaled to the original $1 \times 1$ m$^{2}$ surface element.
Parameters of neutrons reaching the rock/cavern boundary were stored and 
neutrons were propagated later through lead and hydrocarbon shielding. 

\begin{figure}[htb]
\begin{center}
\epsfig{figure=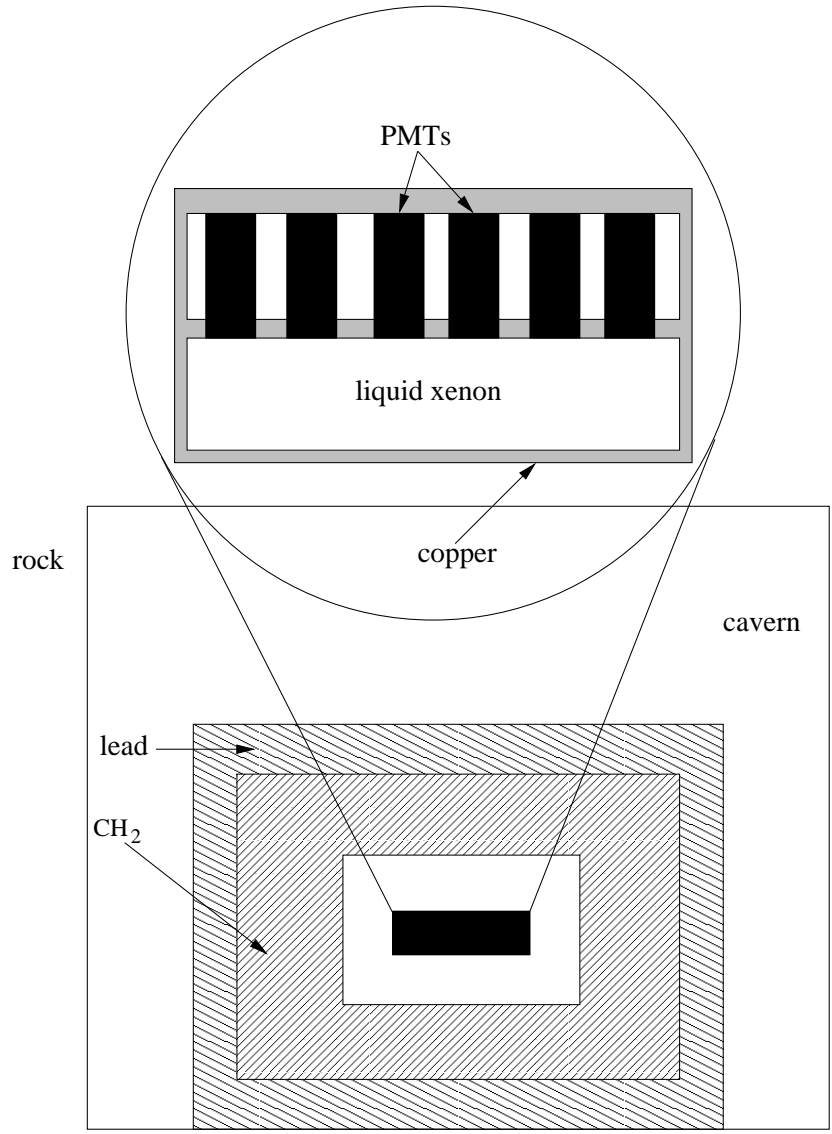,height=15cm}
\caption{Sketch of the laboratory hall with a xenon detector and
shielding inside.} 
\label{fig-det}
\end{center}
\end{figure}

Neutrons were also simulated in a realistic
cavern with a size of 
$30 \times 6.5 \times 4.5$ m$^{3}$ in the rock (see Figure \ref{fig-det}). 
The neutron flux at the rock/cavern boundary for this configuration
was found to be
$4.36 \times 10^{-6}$ cm$^{-2}$ s$^{-1}$ above 100 keV and 
$2.20 \times 10^{-6}$ cm$^{-2}$ s$^{-1}$ above 1 MeV. 
In practice these values are affected by the back-scattering of neutrons
from other walls of the cavern: a neutron can enter the cavern, reach the
opposite wall and be scattered back into the cavern increasing
the total flux through the boundary. 
To check the effect of back-scattering, neutrons were propagated 
through the cavern and counted each time they entered the
cavern. In this case the calculated neutron flux
was $1.19 \times 10^{-5}$ cm$^{-2}$ s$^{-1}$ above 100 keV and 
$4.10 \times 10^{-6}$ cm$^{-2}$ s$^{-1}$ above 1 MeV. 
For a real detector the back-scattering of neutrons
can occur also on detector elements, shielding etc.

\begin{figure}[htb]
\begin{center}
\epsfig{figure=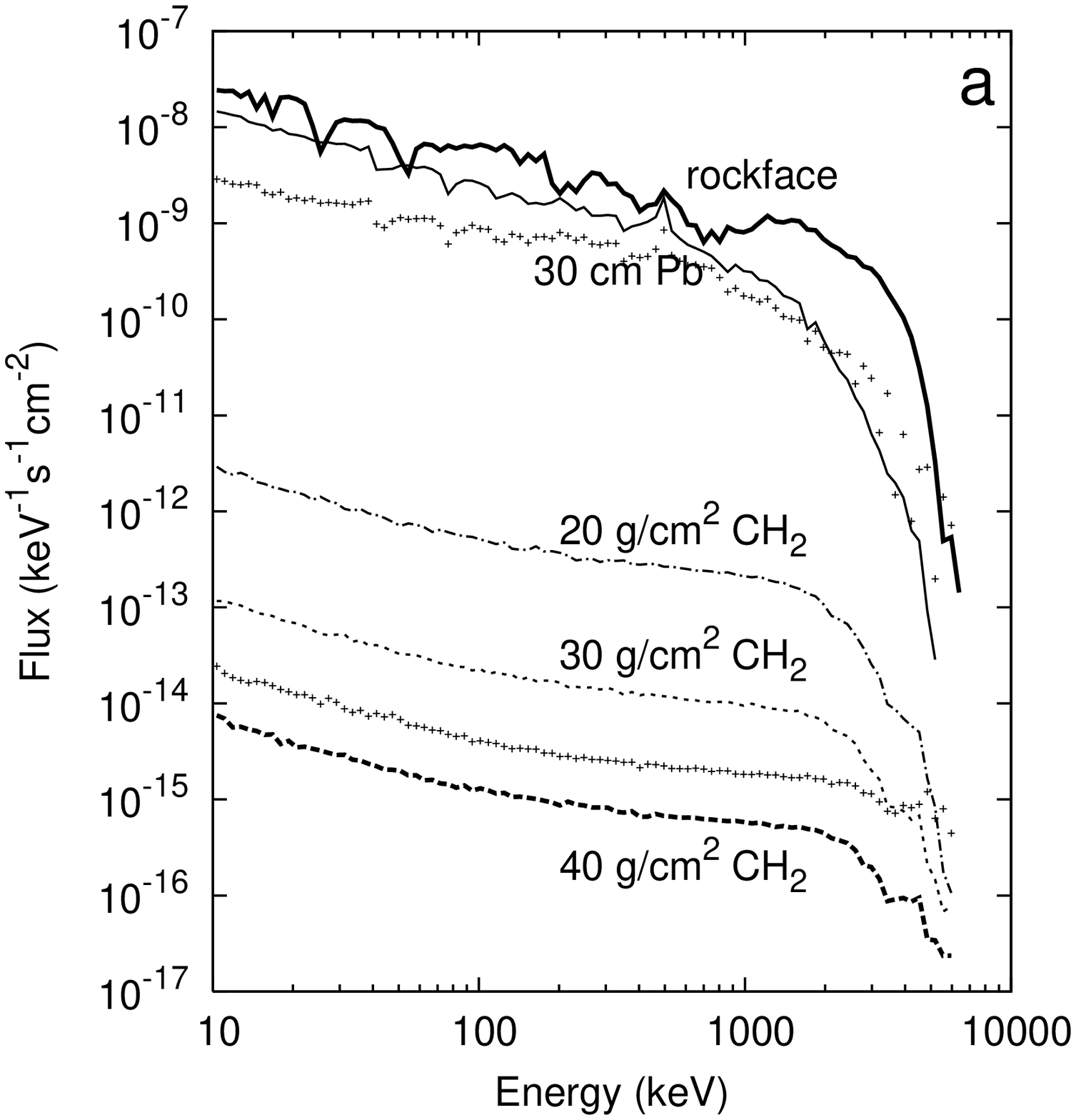,height=7.5cm}
\epsfig{figure=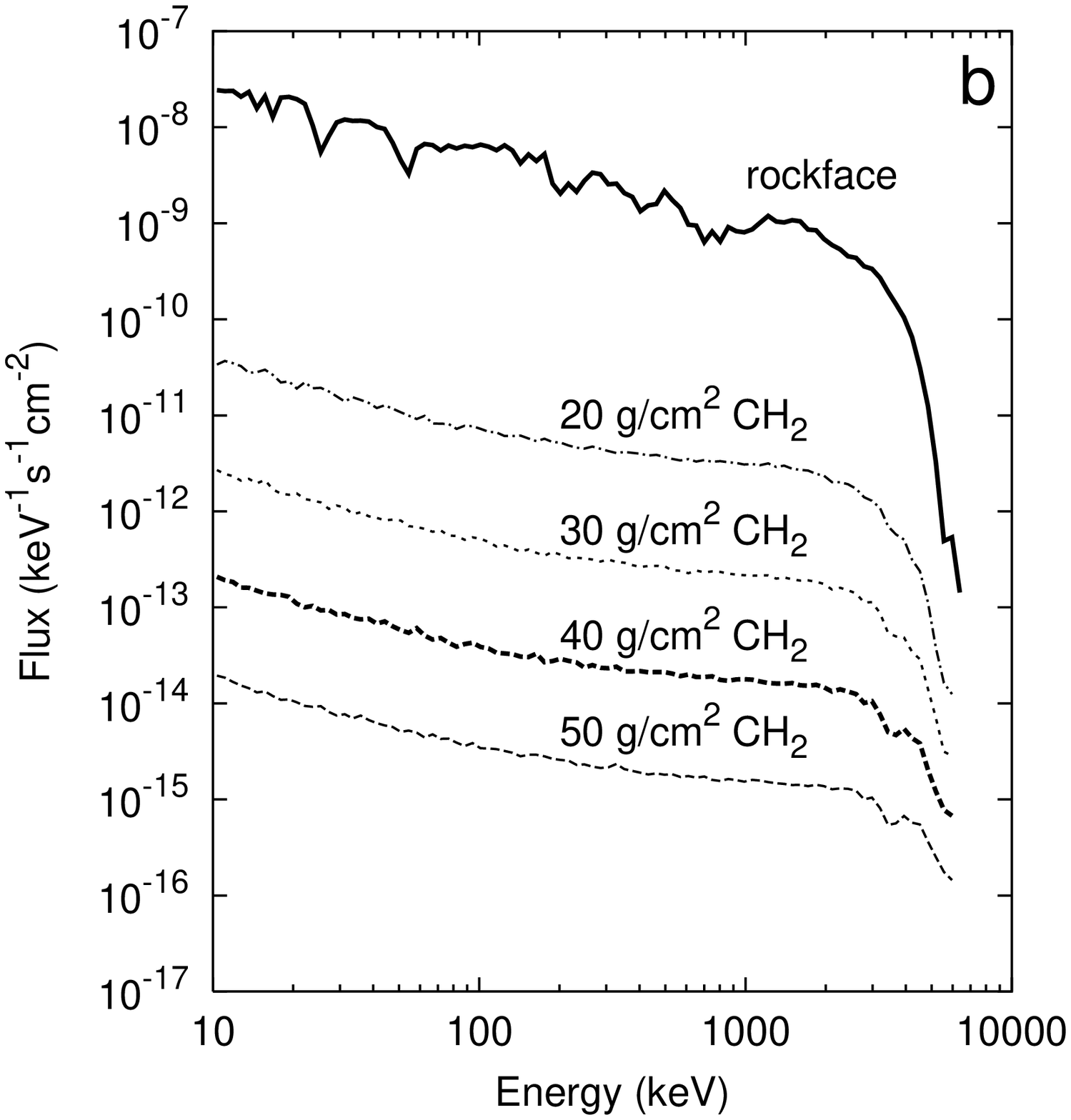,height=7.5cm}
\caption{Neutron energy spectra from rock activity after
lead and hydrocarbon shielding: 
{\it a} -- with 30 cm of lead between 
salt and hydrocarbon; {\it b} -- without lead. Neutrons were propagated
using GEANT4. The initial neutron production spectrum was obtained
with modified SOURCES. 
Thick solid curve -- spectrum at the salt/cavern boundary;
thin solid curve -- spectrum after 30 cm of lead ({\it a});
dash-dotted curve -- spectrum after 20 g/cm$^{2}$ of hydrocarbon shielding
behind lead ({\it a}) or without lead ({\it b});
dotted curve -- spectrum after 30 g/cm$^{2}$ of hydrocarbon;
thick dashed curve -- spectrum after 40 g/cm$^{2}$ of hydrocarbon;
thin dashed curve -- spectrum after 50 g/cm$^{2}$ of hydrocarbon ({\it b}).
Crosses show the spectra obtained with a neutron spectrum
at the rock/cavern boundary from Modane measurements (see text for details):
upper spectrum is for the flux after lead shielding; lower spectrum
is for the flux after lead and 40 g/cm$^2$ of CH$_2$.}
\label{fig-nspch2}
\end{center}
\end{figure}

Lead of low radioactivity is widely used to shield dark matter detectors from 
gammas produced in the rock and the laboratory walls. Some detectors, 
however, can be made insensitive to these gammas \cite{drift}. So the use of 
lead as shielding and its thickness is 
decided for each particular experiment. Normally the thickness is such that 
gammas from rock contribute only a minor part to the total gamma flux 
at the detector, while a major contribution comes from the detector itself.
Neutrons produced via spontaneous fission 
in low activity lead do not contribute significantly to the neutron flux 
coming from the rock.
Simulations were carried out with and without lead shielding to 
investigate the effect of lead on the neutron flux. Neutrons coming 
from the rock in a simple geometry (neutron production volume -- 
$1 \times 1 \times 3$ m$^{3}$, neutron propagation volume --
$100 \times 100 \times 3$ m$^{3}$ as described above)
were propagated through 30 cm of lead and those 
emerging on the opposite side were stored. Note 
that neutrons can be scattered back from the lead into the rock and 
then to the lead again. In order not to lose these neutrons the rock was 
present at this stage of the simulations, although the 
neutrons were generated only on its boundary. Then hydrocarbon 
shielding of varying thicknesses was added to the set-up after lead. 
Similar simulations were carried out without lead (with 
hydrocarbon only). Figure \ref{fig-nspch2}
shows neutron spectra after 30 cm of lead 
and slabs of hydrocarbon of various thicknesses (Figure \ref{fig-nspch2}a), 
and similar spectra obtained without lead (Figure \ref{fig-nspch2}b). 
Due to the high cross-section 
of inelastic neutron scattering in lead above 4 MeV, the neutron 
spectrum after lead is suppressed at these energies. This results 
in a larger suppression of the neutron flux by the hydrocarbon when 
the lead is present compared to the case without lead. 

\begin{figure}[htb]
\begin{center}
\epsfig{figure=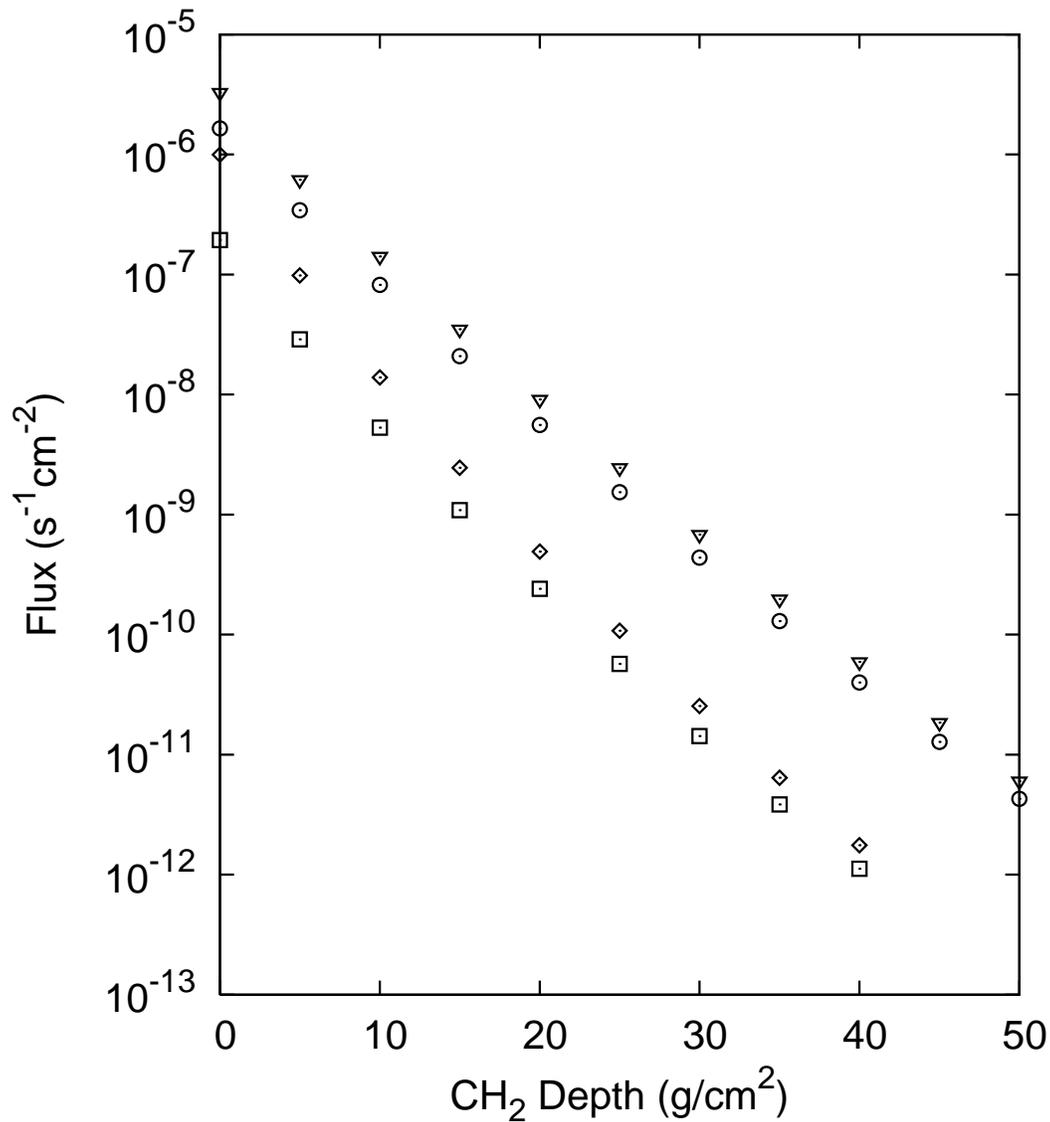,height=15cm}
\caption{Suppression factors for the total neutron flux above 100 keV
and 1 MeV as a function of hydrocarbon thickness with and without
lead shielding. 
Diamonds -- flux above 100 keV beyond 30 cm of lead separating salt 
and hydrocarbon shielding; 
triangles -- flux above 100 keV, no lead between salt 
and hydrocarbon shielding; 
squares -- flux above 1 MeV, beyond 30 cm of lead between salt 
and hydrocarbon shielding;
circles -- flux above 1 MeV, no lead between salt 
and hydrocarbon shielding.}
\label{fig-suppr}
\end{center}
\end{figure}

The change in the neutron flux and spectrum can be expressed in terms 
of a suppression factor, which shows the ratio of the total neutron flux 
above a certain threshold after shielding to the initial flux. 
The suppression factors as functions of the thickness of hydrocarbon 
shielding for neutrons above 100 keV and 1 MeV are shown in Figure 
\ref{fig-suppr}. A lower neutron flux is expected if lead is used in the 
shielding together with the hydrocarbon material.

Present results agree with a preliminary simulation carried out 
with the MCNP code \cite{mcnp}, in which lead, copper and 
hydrocarbon shielding were used \cite{idmneutrons}.

From Figure \ref{fig-suppr} we can conclude that using 35 
g/cm$^{2}$ of hydrocarbon material together with 30 cm of lead or 
50 g/cm$^{2}$ of hydrocarbon material without lead the neutron flux 
can be suppressed by about six orders of magnitude, which is a 
typical requirement to achieve a sensitivity to a WIMP-nucleon cross-section 
of about 10$^{-10}$ pb. Lower (higher) fraction of hydrogen,
compared to our basic CH$_2$ composition, requires
larger (smaller) thickness of hydrocarbon material.

\begin{figure}[htb]
\begin{center}
\epsfig{figure=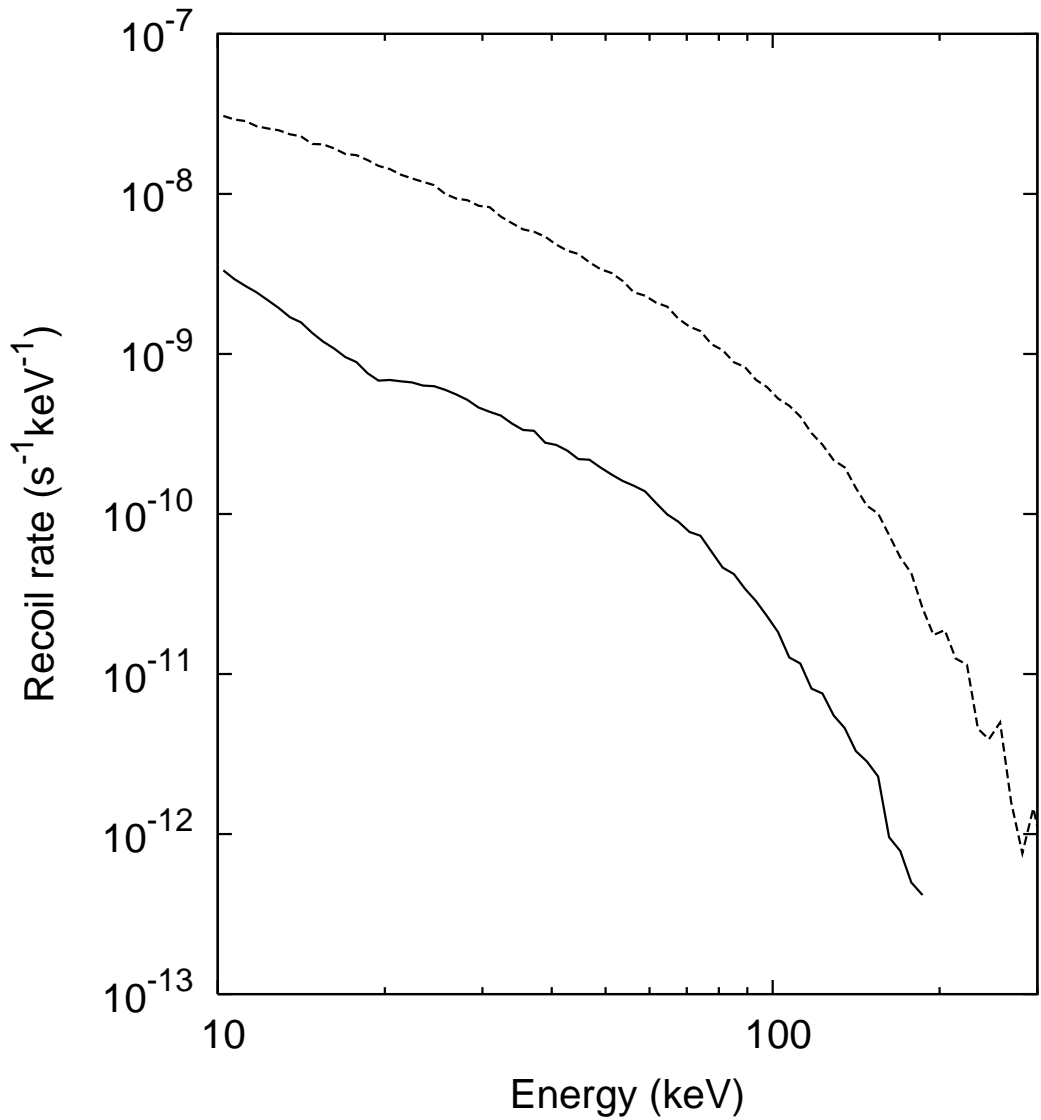,height=15cm}
\caption{Energy spectrum of nuclear recoils in a 250 kg liquid
xenon detector from rock neutrons beyond 30 cm of lead and 35
g/cm$^2$ of hydrocarbon (see text for details).
Dashed curve shows the recoil spectrum originated from
input neutron spectrum from Modane measurements.}
\label{fig-recsprock}
\end{center}
\end{figure}

The precise thickness of hydrocarbon material to be 
used in the shielding depends on the geometry of a detector, target 
material, efficiency of neutron detection and 
the initial neutron flux (radioactive contamination of rock).
Simulations of the nuclear recoil energy spectrum in a 
large-scale xenon detector were carried out using the GEANT4 package 
with a neutron spectrum 
expected for the salt rock. The realistic geometry
of the laboratory hall was taken into account 
(cavern size $30 \times 6.5 \times 4.5$ m$^{3}$).
The neutron flux incident on the shielding around the detector is
2-3 times higher than in a simple geometry due to the back-scattering
of neutrons from the walls.
The energy spectrum of nuclear recoils expected in a 250 kg 
xenon detector (a cylinder with a diameter of 103 cm, height of 10 cm 
and density of 3 g/cm$^{3}$) surrounded by 35 g/cm$^{2}$ of 
hydrocarbon material and 30 cm of lead is shown in Figure 
\ref{fig-recsprock} by a solid line. 
The expected rate in a 10-50 keV recoil 
energy range (2-10 keV electron equivalent energy assuming a quenching
factor of 0.2 for xenon recoils in xenon as reported in 
Ref. \cite{quenching}) is 0.86 events per year.
We conclude that 35-40 g/cm$^{2}$ of hydrocarbon
and 30 cm of lead
are enough to suppress the neutron flux from rock activity down to 
less than 1 event per year in a 250 kg xenon detector. As we will see below, 
this rate is sufficient to reach 
a sensitivity of about 10$^{-10}$ pb to a
WIMP-nucleon cross-section. If no lead is used, then 50 g/cm$^{2}$ is 
needed for a similar nuclear recoil rate in the detector and similar 
sensitivity.

If we assume the contamination levels to be 70 ppb U and 125 ppb Th
(as recently measured in the new cavern at Boulby \cite{ukdmc}),
then the nuclear recoil rate in 250 kg of xenon behind 30 cm of lead 
and 35 g/cm$^{2}$ of hydrocarbon is about 0.56 events per year.

A shield made out of 30 cm of iron instead of lead
suppresses the neutron flux by another order of magnitude after 35-40
g/cm$^2$ of hydrocarbon, but iron is less efficient for gamma
absorption and typically has higher U/Th levels than lead.

To check the uncertainty of the results, we carried out similar simulations
with the modified SOURCES code for Modane rock and compared them to the
measurements of the neutron flux in the Modane underground laboratory 
\cite{modane}. Using the contamination levels for rock from 
Ref. \cite{modane} we obtained a neutron production rate from
($\alpha$,n) reactions of $8.11 \times 10^{-8}$ cm$^{-3}$ s$^{-1}$,
which is about half the value calculated in Ref. \cite{modane},
assuming the rock density of 2.7 g/cm$^3$. The difference is not great
taking into account the uncertainties in the measured and evaluated
cross-sections used, and the difference in the simulation strategy.

\begin{figure}[htb]
\begin{center}
\epsfig{figure=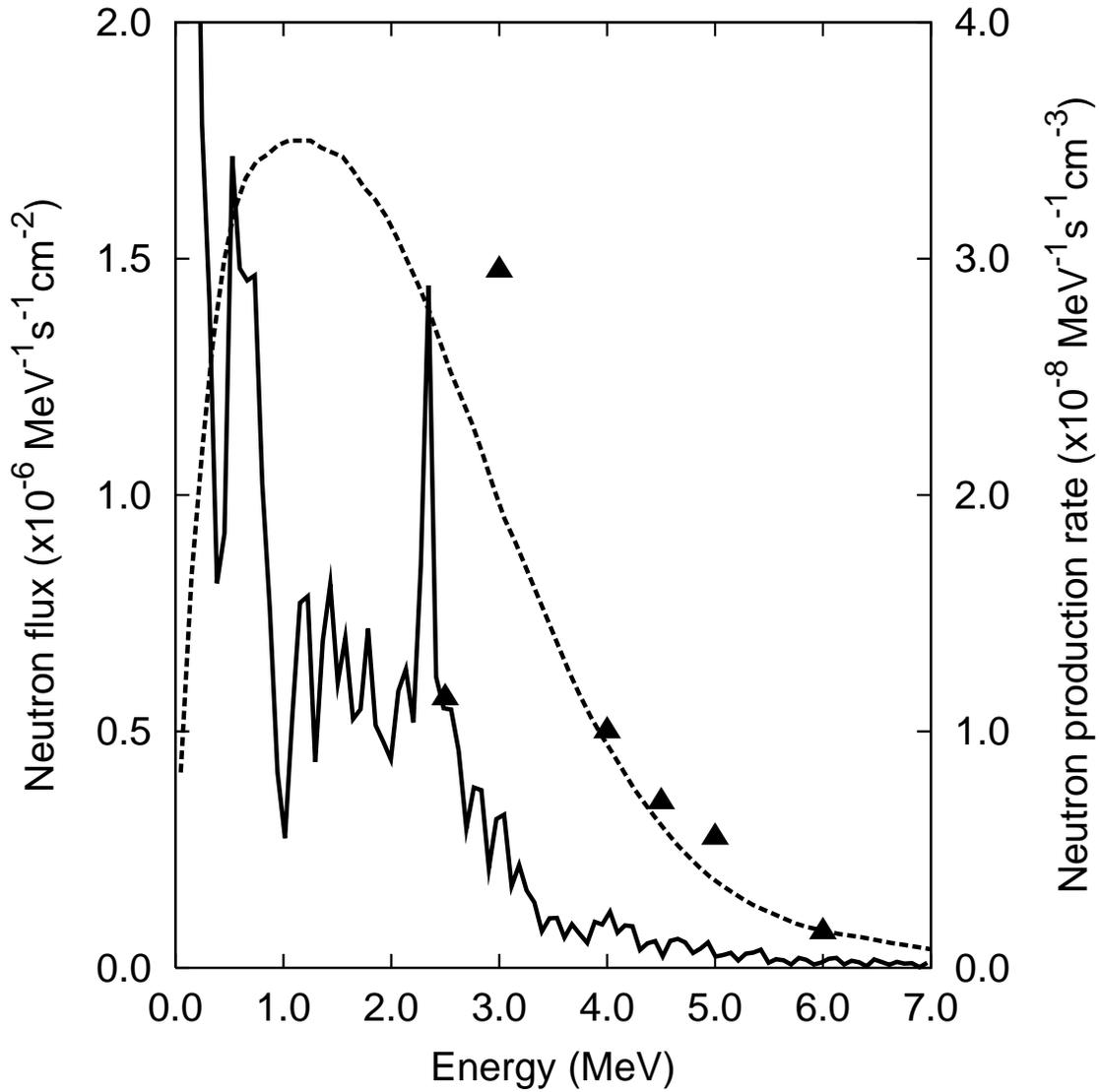,height=15cm}
\caption{Neutron production spectrum 
as calculated with modified SOURCES for Modane rock (dashed curve
and right y-axis for rate units) is plotted together with
the neutron energy spectra at the rock/cavern boundary:
solid curve -- spectrum from SOURCES propagated with GEANT4;
triangles -- evaluated spectrum from measurements
\cite{modane}.}
\label{fig-modane}
\end{center}
\end{figure}

Figure \ref{fig-modane} shows the initial neutron
production spectrum for Modane rock together with the spectra of neutrons
at the rock/cavern boundary calculated with SOURCES (neutron production)
and GEANT4 (neutron propagation), and measured at Modane \cite{modane}.
The two spectra at the boundary look very different. The measured
spectrum has a peak at about 3 MeV, similar to the neutron production
spectrum in rock from SOURCES but shifted to higher energies. The spectrum
of neutrons generated with SOURCES and propagated to the rock surface with
GEANT4 does not have a wide peak (the sharp peak at 2.3 MeV and various 
dips are due to the dip and peaks
in the cross-section of neutron scattering on oxygen). The wide peak
seen in the production spectrum has been smoothed by neutron
interactions in rock. Note that the spectrum obtained in Ref. \cite{modane} 
is not a directly measured neutron spectrum. It was deconvolved from the
measured proton recoil spectrum using a complicated procedure involving
simulations including neutron propagation through the lead and copper 
shielding around the detector. Total neutron flux above 2 MeV has recently
been re-evaluated, using more accurate simulations of the neutron propagation
through the shielding and detector efficiency, and reduced from 
$4.0 \times 10^{-6}$ cm$^{-2}$ s$^{-1}$ above 2 MeV down to 
$1.6 \times 10^{-6}$ cm$^{-2}$ s$^{-1}$ \cite{chardin}. The distortion
of the neutron spectrum after Pb and Cu shielding (underestimated
in Ref. \cite{modane}) was mentioned
as the main cause of the difference in neutron flux \cite{chardin} but
no corrected spectrum was provided. A similar distortion in rock is 
responsible for a change of the neutron spectrum from that at production 
to that at the rock/cavern boundary as our simulations show 
(see Figure \ref{fig-modane}). We believe that, although an obvious
difference between our calculated spectrum and the measured one is observed,
no definite conclusion can be derived about the correct shape of the
neutron spectrum based on the existing data and simulations.

A wide peak (at about 2.2-2.3 MeV) in the measured 
neutron spectrum in the laboratory has also been 
reported recently in Ref. \cite{kim}. Here again the authors
used the deconvolution of the measured spectrum of proton recoils
and no simulation of the neutron propagation through the detector
shielding was performed. No distinctive peaks have been reported for
either of the measurements in the Gran Sasso Laboratory above 1 MeV
\cite{belli,icarus}, although
the evaluated spectra appear to be harder than the spectrum from
our simulations.

Having found a significant difference in the laboratory spectra
(at the rock/cavern boundary) we studied the effect this
may have on the neutron spectrum and nuclear recoil rate behind
the shielding. We propagated neutrons with an energy spectrum 
from Ref. \cite{modane} for Modane rock through the shielding 
in a simplified geometry as described above 
(a slab of rock and slabs of shielding: 30 cm of Pb and 
40 g/cm$^2$ of CH$_2$) and obtained
neutron spectra behind the shielding shown by crosses in Figure
\ref{fig-nspch2}a (upper spectrum is behind 30 cm of Pb, 
lower spectrum is behind lead and  40 g/cm$^2$ of CH$_2$). 
The total neutron flux in the lab above 10 keV
was taken as in our previous simulations to investigate the difference in the
spectrum shape only. The nuclear recoil rate is
proportional to neutron flux and the uncertainty in the total neutron flux
can be easily propagated to the recoil flux.
We assumed isotropic distribution of neutron directions at the rock surface
within a hemisphere.
After lead the 'Modane' spectrum is already not very much different in shape 
to the spectrum originated from SOURCES but is still harder. 
After lead and 40 g/cm$^2$ of CH$_2$ the 'Modane' spectrum below 1 MeV
has similar shape to the spectrum originated from
SOURCES but the neutron flux is three times higher. Above 1 MeV
the difference in neutron flux is larger.

The effect is more dramatic if a realistic geometry of the cavern, shielding
and detector is taken into account together with the production of
nuclear recoils. We propagated neutrons from
the rock through the cavern and shielding around the detector and
calculated the spectrum of nuclear recoils in a detector.
The cavern had a size of $30 \times 6.5 \times 4.5$ m$^{3}$,
the detector was a cylinder with a diameter of 103 cm, height of 10 cm 
and density of 3 g/cm$^{3}$ surrounded by 35 g/cm$^{2}$ of 
hydrocarbon material and 30 cm of lead, as in our original simulations
for recoil spectrum in xenon.
The results are plotted in Figure \ref{fig-recsprock} (dashed curve). 
The spectrum is similar to the recoil spectrum originated from SOURCES, 
but the recoil rate in 10-50 keV energy range is 16 times higher 
(13.6 events/year). So, for a recoil flux in a realistic geometry, we
obtained larger difference between our
original spectrum and 'Modane' spectrum (Figure \ref{fig-recsprock})
than for neutron flux in a simplified geometry (Figure
\ref{fig-nspch2}a). This large difference comes from the
harder 'Modane' spectrum. 

We conclude that, i) realistic geometry of the cavern, shielding and 
detector is important for evaluation of the neutron suppression and
nuclear recoil rate; ii) a much harder neutron spectrum results
in a significant increase in the recoil rate (in our case the 
'Modane'-type spectrum requires an additional 10-15 g/cm$^2$ of CH$_2$).

Special attention should be paid to the background associated with 
radon. Due to very high permeability of radon, it can penetrate 
through materials and contribute to 
the background in the target. Alpha background is briefly 
discussed in Section 6. 
Radon or its daughter's decay can occur in xenon close to the 
vessel wall or in the wall close to xenon with an alpha going out of
the detector and a recoiling nucleus going into the detector and
contributing to the low energy nuclear recoil rate. This effect is similar
to that previously seen in some NaI(Tl) \cite{bump,gilles}
detectors. These events can easily be rejected in a large xenon 
detector with position sensitivity.

Alphas from radon decay can also 
add to the neutron background in the vicinity of the detector, if radon 
penetrates through the shielding via the air gaps (large thickness of lead
and hydrocarbon makes the diffusion of radon through the shielding 
unlikely). 
Materials surrounding the xenon target consist mainly of isotopes with 
a high threshold for ($\alpha$,n) reactions, such as copper, 
iron, nitrogen and oxygen in air
and carbon in hydrocarbon shielding. As a consequence of this, the 
neutron background from U/Th in 
these materials is dominated by spontaneous fission of 
U$^{238}$, with ($\alpha$,n) reactions contributing about 20\%. 
Radon and its daughters are free from spontaneous fission and only 
($\alpha$,n) reactions are the source of neutrons from radon decay.
The main threat from radon is that if a significant fraction of rock
produced radon is allowed to penetrate the shields, the alpha emitting
progeny can be deposited close to the detector.
Assuming a radon decay rate of about 1 Bq/m$^3$ -- a typical value
for a ventilated laboratory, the neutron flux from alphas from radon
decay can be a significant fraction of the flux from 
ultra-low-background PMTs discussed
in Section 5. To suppress this flux, an appropriate 
protection against radon should be used, such as gas-tight sealing.

\vspace{0.5cm}
{\large \bf 4. Muon-induced neutrons}
\vspace{0.3cm}

Simulations of neutron production by muons with various energies
in hydrocarbon material were recently performed by 
Wang et al. \cite{wang} using the FLUKA Monte Carlo code \cite{fluka}.
Kudryavtsev et al. \cite{vakneutrons} studied neutron production
by muons in various materials with FLUKA using a calculated spectrum and 
an angular distribution of muons at the Boulby Underground Laboratory.
This study is extended in the present work by including
shielding materials and the production of nuclear recoils in the xenon
detector.

The muon spectrum and angular distribution was simulated using the MUSUN 
Monte Carlo code (see Ref. \cite{vakneutrons} for description). 
Normalisation of the muon (and neutron) spectrum was done using the 
measured value for the muon flux at the Boulby Underground Laboratory: 
$(4.09 \pm 0.15) \times 10^{-8}$ cm$^{-2}$ s$^{-1}$, 
which corresponds to a rock overburden at vertical of 
$2805 \pm 45$ m w.~e. \cite{idmneutrons,muflux}. The input surface 
muon spectrum was taken in the form suggested by Gaisser 
\cite{gaisser} with the parameters from the best fit to the LVD 
underground muon data \cite{lvd}.
For any other experimental site at similar depth, the neutron flux is 
scaled roughly as the muon flux. Variation of muon energy spectrum
with depth can be accounted for by using the dependence of neutron
production on the mean muon energy as $\propto E^{0.79}$ \cite{vakneutrons}
(or $\propto E^{0.74}$ \cite{wang}). 
Changes in the neutron flux due to different rock composition 
around the laboratory (pure NaCl
was used in this work) can be estimated following the dependence
of the neutron production on the mean atomic weight of the rock as
$\propto A^{0.76}$ \cite{vakneutrons}.

Muons were sampled on the surface of a cube of rock (NaCl)
$20 \times 20 \times 20$ m$^{3}$. The laboratory cavern of size 
$6 \times 6 \times 5$ m$^{3}$ was placed inside the salt region
at a depth of 10 m from the top of the cube and at a distance of 7 m 
from each vertical surface of the cube. The cavern contained shielding
made of lead and hydrocarbon material (see Figure \ref{fig-det}). 
Typical thickness was 30 cm of 
lead and 40 g/cm$^2$ of hydrocarbon material, corresponding to 
about 45-50 cm of liquid scintillator (active shielding) or about 
40-45 cm of wax or polyethylene. The hydrocarbon was placed
inside the lead so that it would absorb neutrons produced by muons in the 
lead. As will
be shown below this is very important for shielding against
muon-induced neutrons. A cylindrical vessel (103 cm diameter, 10 cm
height) made of stainless steel of a thickness of 2 cm
containing 250 kg of liquid xenon (density 3 g/cm$^3$) was placed 
inside the shield. As iron and
copper have similar atomic weights, substitution of iron with copper 
in a detector vessel will not change the neutron production in the 
vessel.

Simulations of muon propagation and interactions, development
of muon-induced cascades, neutron production, propagation and 
detection were performed with FLUKA \cite{fluka}. Tests of 
muon-induced neutron simulations with FLUKA can be found in 
\cite{wang,vakneutrons}. 
The neutron production rate in NaCl at Boulby was found to be 
$7.6 \times 10^{-4}$
neutrons per muon per 1 g/cm$^2$ of muon path. The total neutron flux
at the salt/cavern boundary is $8.7 \times 10^{-10}$ cm$^{-2}$ s$^{-1}$
above 1 MeV.

\begin{figure}[htb]
\begin{center}
\epsfig{figure=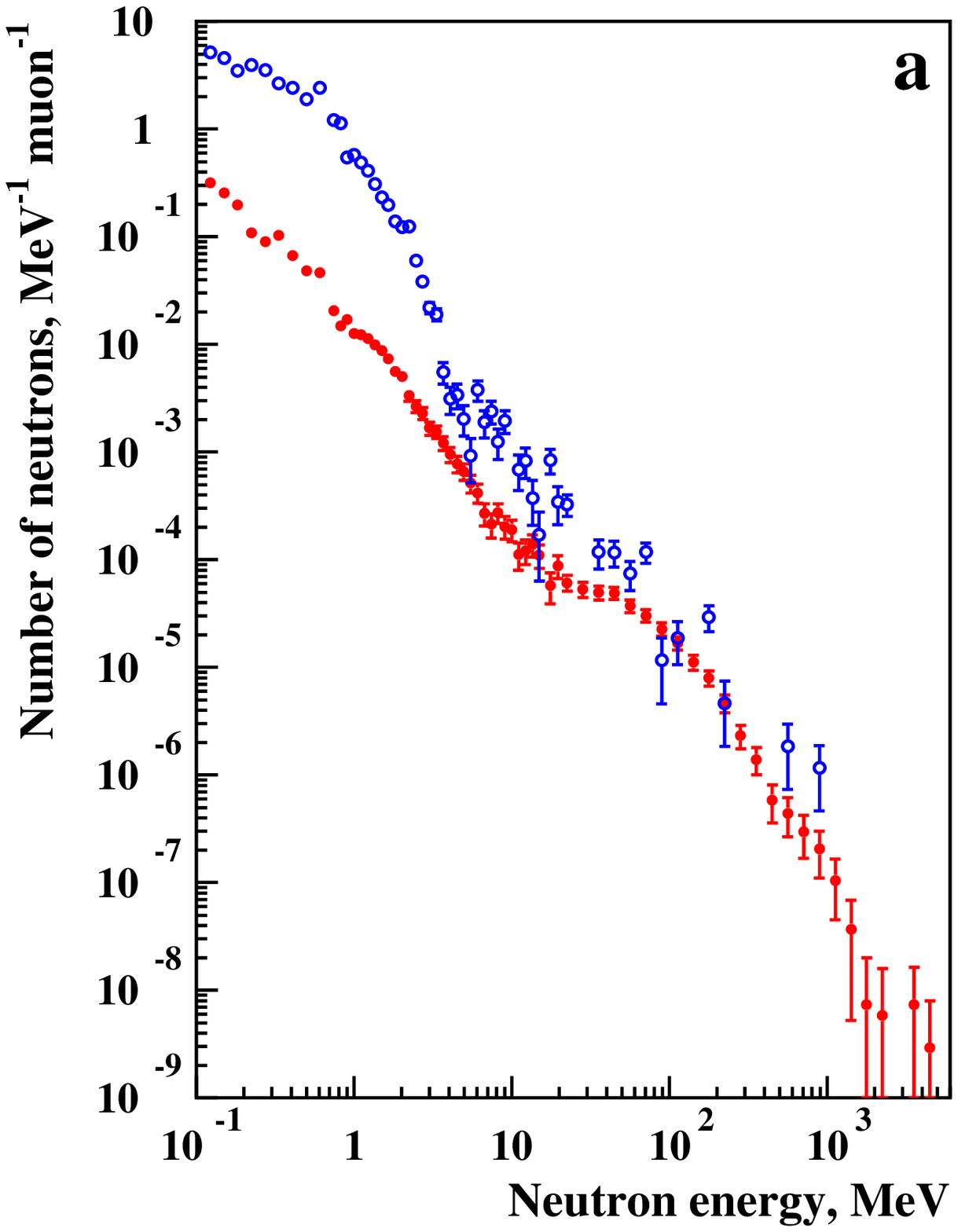,height=9cm}
\epsfig{figure=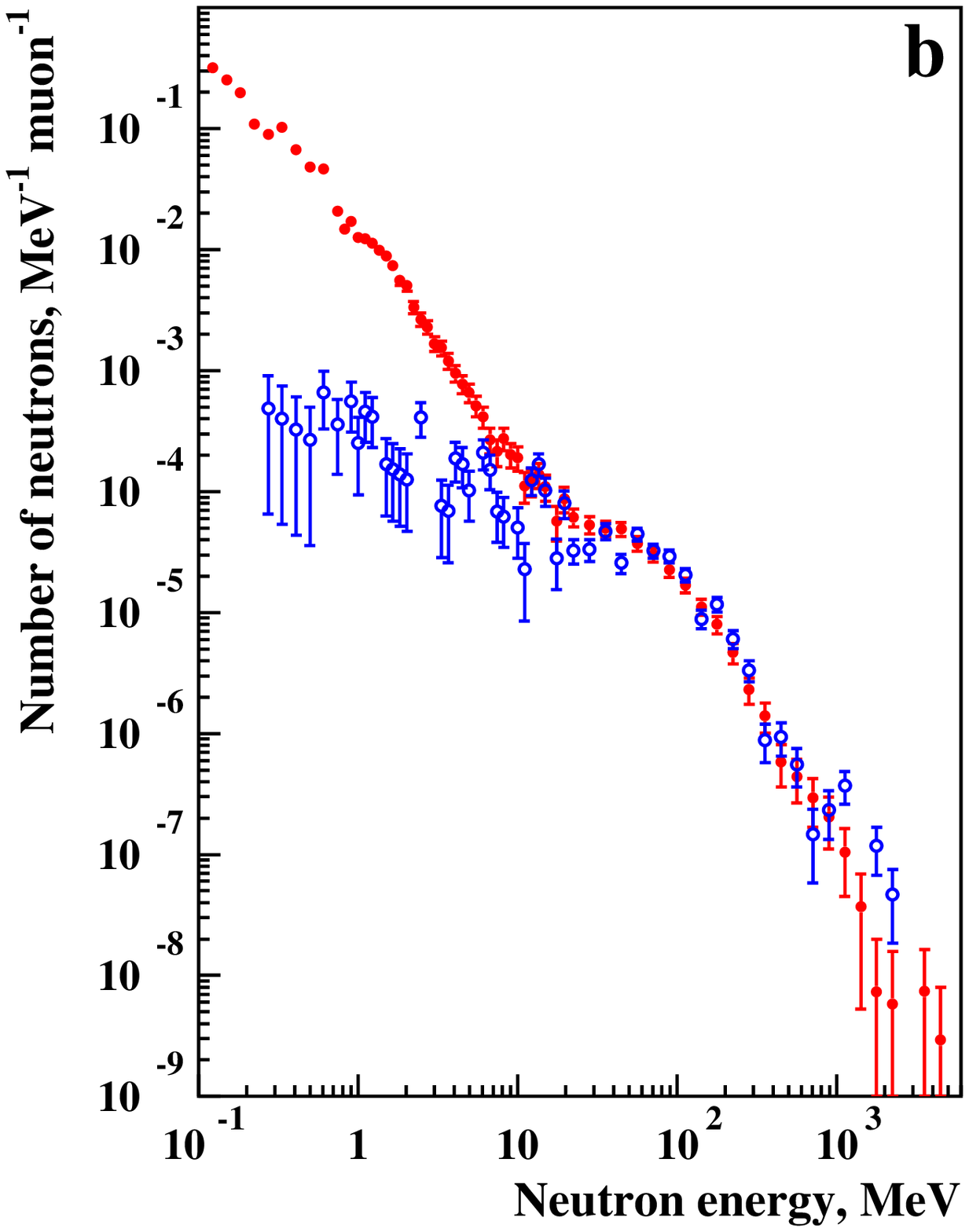,height=9cm}
\caption{Energy spectra of muon-induced neutrons
at various boundaries: 
{\it a} -- filled circles - neutrons at the salt/cavern boundary,
open circles - neutrons after the lead shielding;
{\it b} -- filled circles - neutrons at the salt/cavern boundary 
(the same as in Figure {\it a}),
open circles - neutrons after the lead and hydrocarbon 
shielding.} 
\label{fig-nsp}
\end{center}
\end{figure}

Figure \ref{fig-nsp} shows the effects
of lead and hydrocarbon material on neutron production and
absorption. Neutron spectra on the rock/cavern boundary and
after the lead (lead/cavern boundary) are presented in 
Figure \ref{fig-nsp}a. 
The large increase in the neutron spectrum after the lead,
in particular below 1 MeV, is due to efficient neutron
production in lead. Hydrocarbon material suppresses this flux
by a large factor at low energies (Figure \ref{fig-nsp}b).
This figure demonstrates the potential danger of placing the lead
shielding close to the detector and inside the hydrocarbon 
shielding. Active veto systems may help to reject 
many events associated with muons but they cannot be 100\% 
efficient if they are placed around the main detector and the 
lead shielding. The neutron flux is suppressed by a factor
of $10^2-10^4$ by 40 g/cm$^2$ of hydrocarbon shielding at
energies 0.1-10 MeV -- most important for low-energy nuclear 
recoil production in xenon. To achieve similar suppression with
active veto systems, they would need to have an efficiency up to 0.9999.
This is very difficult to reach in practice, especially 
if the veto system is large and made of several modules.
Another possibility is to use a single module veto detector placed 
just around the target. Such a detector, being made of 
hydrocarbon scintillator, can substitute for the passive absorber and
provide additional rejection capabilities, which will be discussed
later.

Neutrons produced in and around the detector can give nuclear
recoils, which mimic WIMP-induced signals. Nuclear recoils
produced via elastic scattering are of primary interest, since
they are not accompanied by gammas and cannot be discriminated
from WIMP signals. Since many particles in addition to neutrons 
can give an energy deposition in a detector in any particular event
(neutrons are produced mainly in muon-induced cascades), all
of them should be followed to and through the detector with the same
Monte Carlo code. Unfortunately, FLUKA does not generate nuclear
recoils below 19.6 MeV neutron energy realistically. 
Kerma factors, equivalent to the
average energy deposition, are calculated for neutron interactions.
We modified this by assigning to the nuclear recoil an energy equal
to the difference between initial and final neutron energies, which
is correct for neutron elastic scattering. Since FLUKA uses a
multigroup approach for low energy neutrons (below 19.6 MeV), the
recoil energy is different from zero only if a neutron moves from one
group to another one as a result of scattering. If this is not the
case, then the nuclear recoil energy is taken from the initial FLUKA
calculation of kerma factors. Such an approach is reasonable if
there are several neutron interactions in the target and energies of
recoils are summed, decreasing the uncertainty related to a single
interaction. For single recoil
events the statistical approach should be used with caution and
the uncertainties associated with nuclear recoil treatment should be
investigated in more detail.

\begin{figure}[htb]
\begin{center}
\epsfig{figure=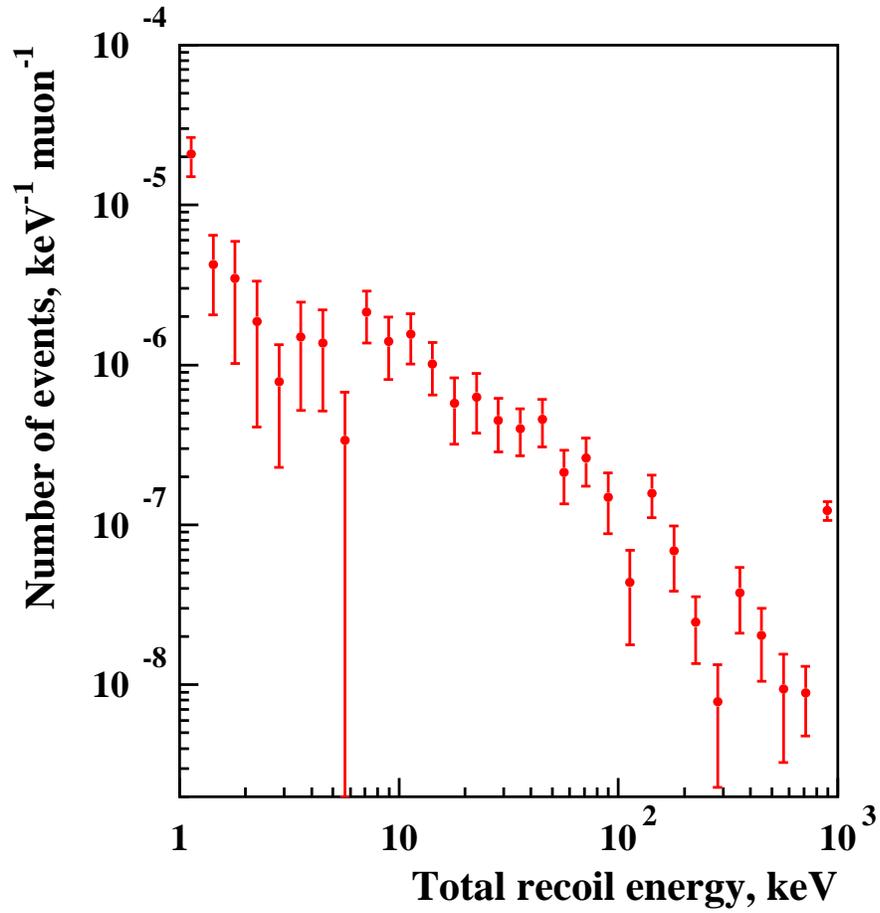,height=15cm}
\caption{Recoil energy spectrum in a 250 kg liquid xenon
detector from muon-induced neutrons. 
The energy of all recoils in any particular event
was summed to give the energy of each event. The highest energy
point plotted also includes all events above 1 MeV.} 
\label{fig-recsp}
\end{center}
\end{figure}

\begin{figure}[htb]
\begin{center}
\epsfig{figure=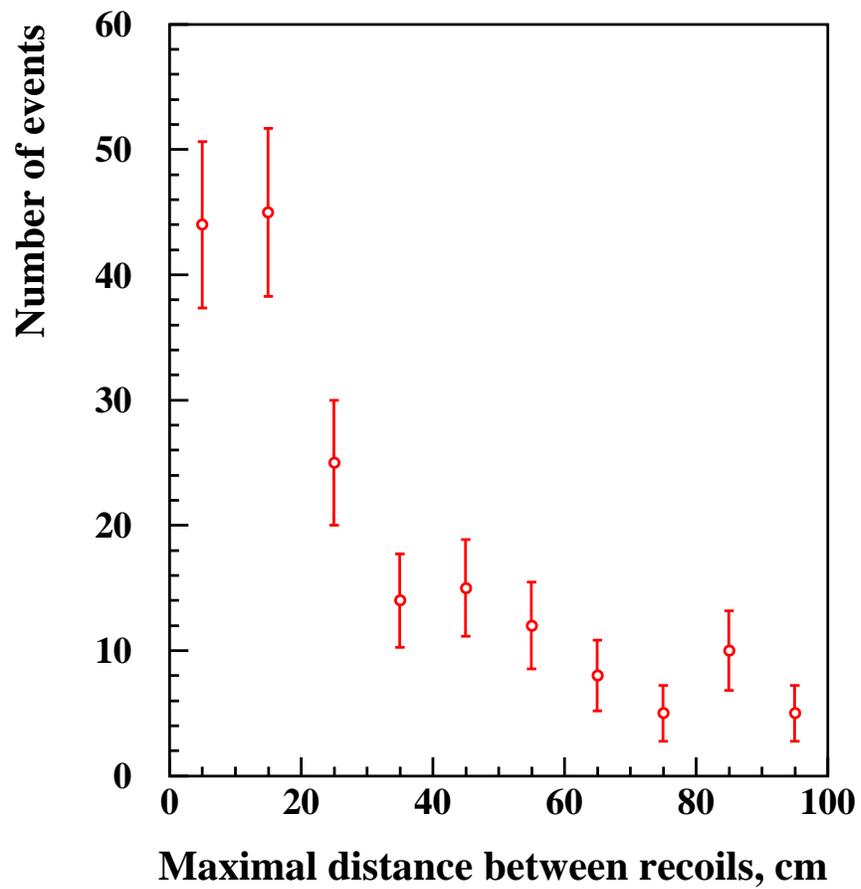,height=15cm}
\caption{Distribution of maximal distances between
nuclear recoils in events occuring in a 250 kg liquid xenon
detector from muon-induced neutrons.} 
\label{fig-recdist}
\end{center}
\end{figure}

Figure \ref{fig-recsp} shows the energy spectrum of nuclear recoil
events originated from muon-induced neutrons 
in a 250 kg liquid xenon detector. Events with multiple recoils
were assigned an energy equal to the sum of the individual nuclear recoil
energies, but excluding energy depositions due to processes not
associated with neutron elastic scattering. The maximal distance between
nuclear recoils in multiple recoil events is presented in Figure
\ref{fig-recdist}. A significant fraction of the multiple recoil events
have recoils separated by more than 10 cm.
Rejection of multiple recoil events, which cannot be caused by WIMPs,
should therefore be possible in liquid xenon detectors sensitive
to the recoil position with sufficient accuracy.

About 20 million muons were simulated in total, which corresponds
to a live time of 2.8 years. A total of 250 nuclear recoil
events per year is expected in a 250 kg xenon detector. This also includes
events where nuclear recoils are in coincidence with any other form of
energy deposition in the detector
not associated with nuclear recoils (electrons, photons, muons etc.). 
A rate of 25 events per year corresponds to
nuclear recoils only without any other energy deposition. This is
the number that determines the sensitivity of the detector to
WIMPs, since other events will be rejected as electron-like events.
Out of this, 11 events per year are single nuclear recoils. Finally, 7.5 
nuclear recoil events per year (without electron-like component)
are expected to be within an energy range of interest for dark matter
searches (10-50 keV recoil energy or 2-10 keV measured energy with
a quenching factor of 0.2 \cite{quenching}). Less than a half of them
are single recoils.
This rate has an uncertainty associated with the treatment 
of nuclear recoil energy in FLUKA.

We also considered the possibility of using a
scintillator made out of hydrocarbon material around the target
as an active veto system (single module detector) against 
muon-induced events. 
To check the efficiency of such a veto
only those events detected in anticoincidence between the main xenon
detector and scintillator veto made of 40 g/cm$^2$ thick
hydrocarbon were recorded. No events occuring in the target
were detected in anticoincidence with signals in a hydrocarbon 
scintillator with an energy threshold of 100 keV.

Cosmic-ray muons and their secondaries will also add to the 
electron-like background in xenon. These events will be seen as high 
energy depositions and will be easily discriminated from expected 
nuclear recoil signal. Another problem is connected to the activation 
of some isotopes in detector components, including xenon isotopes, 
by cosmic rays (mainly neutrons) resulting 
in delayed (with respect to a muon) signals. This induced background, 
however, will contribute only a small part to the total gamma 
background from local radioactivity, which is expected to be rejected 
by discrimination. 

Detector components, including xenon, 
can be activated at the surface by much higher cosmic-ray flux, prior 
to moving detector parts underground. 
Fortunately, xenon does not have long-lived radioactive isotopes and
the background from xenon activation should not be a problem.
Using powerful neutron sources 
underground for detector calibration may cause further activation of 
detector components.
Activated isotopes will contribute to the expected
electron-like background, which requires an accurate study in 
connection with a projected discrimination power of a future detector.
The discussion of the discrimination power of a 
xenon dark matter detector is beyond the scope of this paper.

\vspace{0.5cm}
{\large \bf 5. Neutrons from detector components}
\vspace{0.3cm}

This is probably the dominant source of background, and is certainly
the most difficult to calculate. It can come from the readout system,
target, vessel walls, shielding, support structure etc.; 
for all these components the
actual contamination levels are not known precisely. Figures for contamination
levels supplied by manufacturers are usually approximate and can differ
significantly from sample to sample. Measurements of contamination
cannot be done for absolutely all components and again show quite
large variations between samples. An example of this
is the U/Th traces in PMT components. Here we used typical contamination
figures provided by manufacturers together with measurements carried out
by various experimental groups 
(see the UKDMC web-site \cite{ukdmc} for references).

We considered here in detail photomultiplier tubes as probably 
the most important source of background
for a liquid xenon detector. We assumed a detector made out of copper,
for which U/Th levels are well below 1 ppb and do not
pose a serious threat in terms of neutron background (also because
of the high threshold of ($\alpha$,n) reactions in copper).
We will discuss this in more detail later on in this Section.
Concentrations of U and Th in the hydrocarbon material of the
shielding may be slightly higher, but ($\alpha$,n) reactions can occur 
only on $^{13}$C (due to the high threshold of reaction on $^{12}$C),
which is only 1.1\% of the carbon isotopic composition.
Moreover, hydrocarbon is very efficient at slowing down sub-MeV neutrons
and most of it is further away from the xenon target than many other
components. If used as an active veto system, hydrocarbon could
provide a coincidence signal: proton recoil from neutron
scattering and/or thermal neutron capture by a proton or
by an additive element with a large neutron capture cross-section, 
for example Gd. 
Xenon itself can be purified to reduce the neutron background from
spontaneous fission to a very low level.
Remaining neutron-induced events can be discriminated using
energy deposition of fission products, which will occur at a location
separated from neutron-induced recoil.
These considerations show that with the standard PMT readout, 
the PMTs and their bases (including high voltage dividers) should constitute 
the most serious limitations to the detector sensitivity.

Two types of PMTs were used to simulate PMT related neutron background.
The 2-inch ETL type 9266 PMT (the actual diameter of the PMT is 5.2 cm)
with low contamination levels
of U and Th \cite{etl} (see Table \ref{table-ncomp} for details)
was modelled as a cylinder made out of glass, metals and ceramics.
Similarly, the newly designed 2-inch Hamamatsu R8778 tube \cite{hamamatsu} 
(the actual diameter of the PMT is 5.7 cm)
was modelled as a cylinder made out of quartz glass and metals.
A cylindrical
xenon detector, similar to those described in Section 5 
(see Figure \ref{fig-det}), was viewed
by either 217 ETL 9266 PMTs or 169 R8778 PMTs. The total area of the detector
covered by the PMTs is similar in both cases.
The hydrocarbon material is placed between PMTs to reduce neutron flux. 
Note that the window in the standard version of the ETL 9266 PMT is made
out of borosilicate glass and cannot be used for detection of VUV light from xenon.
The solution to this problem, 
which does not lead to an increase in contamination
due to the use of impure materials, is to coat the window with a
wavelength shifter.
The contamination levels for the ETL 9266 PMT were taken from \cite{etl}.
Radioactive impurities in the R8778 PMTs were estimated based
on measurements of activity concentrations from Ref. \cite{hamamatsu}.
Different measurements, however, lead to different contamination
levels ranging from about 2 ppb to 10 ppb of U and Th for
the metal PMT of weight 160 g.
We assumed the contamination levels to be 4 ppb U and 4 ppb Th.
We estimated the proportion of glass and metals in R8778 from similar figures
for other PMTs of similar size. Note that the weight of metal in the R8778
as well as its total weight is larger than that for the ETL 9266 because
the whole envelope of the R8778 is made out of metal (we assumed 
everywhere that the metal of the PMT is stainless steel).

The modified SOURCES code (see Section 2 for details) was used to
obtain neutron spectra from the various materials.
The material component of each PMT was populated separately with the
neutron spectrum from SOURCES and this spectrum was then
emitted isotropically and propagated through the detector.
The setup included 1 cm thick copper walls around the liquid xenon and
PMTs, a 1 cm thick copper support structure for the PMTs and an acrylic
(plexiglass or PMMA) absorber between PMTs to suppress the neutron flux
(see Figure \ref{fig-det}).
If a neutron scattered two or more times in xenon target, 
the energies of all recoils were added up to obtain the total 
measured energy of the event.
In the calculation of recoil rates and sensitivities
we assumed again that a detector had a step function 
energy threshold of 2 keV (10 keV recoil energy with a 
quenching factor of 0.2) and we used a 2-10 keV measured energy range.
The statistics for all simulations 
described in this Section 
correspond to a live time of more than 1000 years of running.

\begin{figure}[htb]
\begin{center}
\epsfig{figure=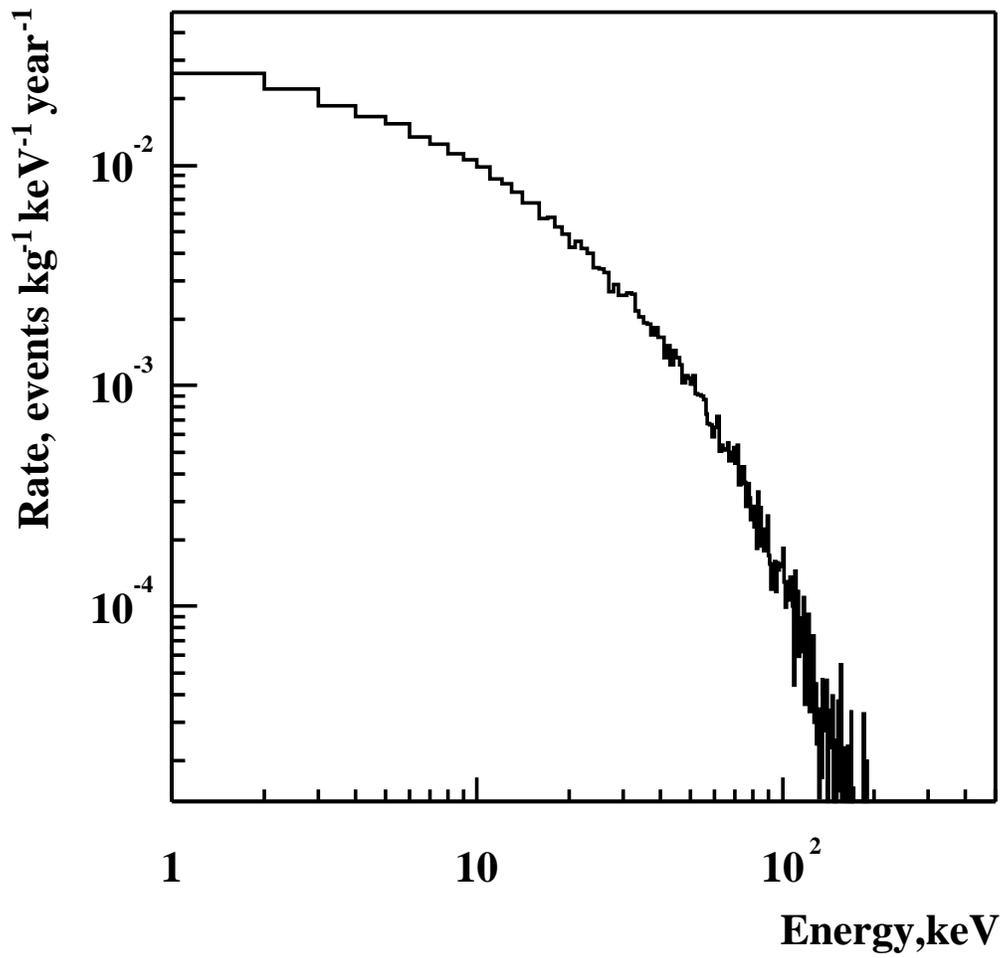,height=15cm}
\caption{Energy spectrum of nuclear recoils produced by neutrons from 
R8778 PMTs in a 250 kg liquid xenon detector.} 
\label{fig-ensp}
\end{center}
\end{figure}

Table \ref{table-ncomp} summarizes the contamination
levels from PMT components \cite{etl,hamamatsu} and the 
simulation results.
Statistical errors for all figures in the table do not exceed 5\%.
The total error is dominated by systematics related to
uncertainty in the contamination levels of detector components.
Such an uncertainty can be as high as a factor of 2.
About 300 events per year at 10-50 keV recoil energies 
are expected in a 250 kg xenon
detector from 217 ETL 9266 PMTs covering about a half of the xenon surface. 
(Note that this number is more than an order
of magnitude higher if standard, not low background, PMTs are used,
which is not acceptable for a sensitive detector.)
Only 7.6 neutrons per year can be detected with 169 R8778 tubes.
Figure \ref{fig-ensp} shows the energy spectrum of
nuclear recoils produced by neutrons from the R8778 PMTs.

Several methods can be used to reduce these numbers.
5 cm thick acrylic (plexiglass) lightguides 
between PMTs and xenon can reduce the nuclear recoil rate
from PMTs by a factor of 2, giving 3.6 events per year from all R8778 PMTs. 
Acrylic is known to have a very low
level of impurities. Contaminations of U and Th at a level of 
$10^{-12}$ ppb have been achieved by the SNO Collaboration \cite{sno}, so
the neutron flux from the lightguides can be negligibly small.
However, the use of the lightguides reduces the light collection.
Moreover, acrylic is not transparent to VUV radiation and
therefore a waveshifter coating for lightguides is needed.

Reducing the contamination levels of PMTs and
making these PMTs of larger (5-inch) size will improve
the sensitivity of future detectors.
Contamination levels of 1 ppb for both U and Th can
reduce the neutron background rate in the detector from PMTs to
0.9 events per year. The use of PMTs with a larger diameter
photocathode (5-inch) and similar impurity levels can further reduce
the nuclear recoil rate bringing it down to about 0.4 events per year,
by reducing the total weight of PMT material.

Let us now consider other sources of neutrons in the detector.
High voltage dividers required by PMTs, with standard resistors and capacitors,
can have high background radioactivity. 
Although the mass of ceramics used in them is small,
the level of impurities reaches 100-1000 ppb of U and Th, giving a rate
potentially comparable to that expected from the PMTs themselves. A way
around this is to use chip resistors and film capacitors with lower
levels of radioactive impurity and to move them further from the target
\cite{hamamatsu}, bringing the number of events from this source
below the PMT contribution. Note that use of the hydrocarbon between PMTs
and as lightguides will also reduce the number of events from 
dividers.

Although the copper vessel is assumed to be heavy (1 cm thick walls were
used in the simulations giving the total weight of copper of 350 kg), 
it should not give a major contribution to the neutron background. 
The radioactive impurities in some copper samples were found to be
below 0.02 ppb \cite{ukdmc}. With such a level of impurity (0.02 ppb) the
nuclear recoil rate is about 0.4 events per year, which can be slightly 
reduced by PTFE (teflon) reflectors, which may be required 
between the copper vessel and xenon target.
PTFE itself contains less than 1 ppb of U and Th and its weight
can be reduced to bring the neutron rate to below 1 event per year.

Further, more precise measurements of the
radioactive impurities in various material samples used in the
detector construction are needed.

Adding up contributions from the main detector components
we can conclude that the total rate of neutron-induced nuclear recoils,
achievable at the present level of technology, 
is of the order of 4-8 events per year in 250 kg of liquid xenon
(10-50 keV recoil energy). With provisional success
in PMT developments and the use of other ultra-pure materials
(for instance, ultra-pure copper) the rate can be reduced
to about 1 event per year (see Table \ref{table-sum}).

Similar or even larger reduction in the background could be achieved with
new readout designs, such as GEM \cite{gem} or MICROMEGAS 
\cite{micromegas}, without PMTs and associated background,
currently being studied for large dark matter detectors 
\cite{buz,ben}.
Materials, such as copper, kapton and teflon, all with very
low levels of impurities can be used reducing the number of expected
events to less than 1 event per year (see Table \ref{table-sum},
Detector 5 described in Section 6).

Another possibility to suppress the background is to estimate it
by independent techniques and subtract the estimated value from
the measurements. In practice, this can be done using
statistical tables \cite{pdg} for the upper limits on a signal
at a certain confidence level as functions of the measured number 
of events and estimated background (neutron-induced nuclear recoils).
The following methods to estimate the background can be used.

\begin{figure}[htb]
\begin{center}
\epsfig{figure=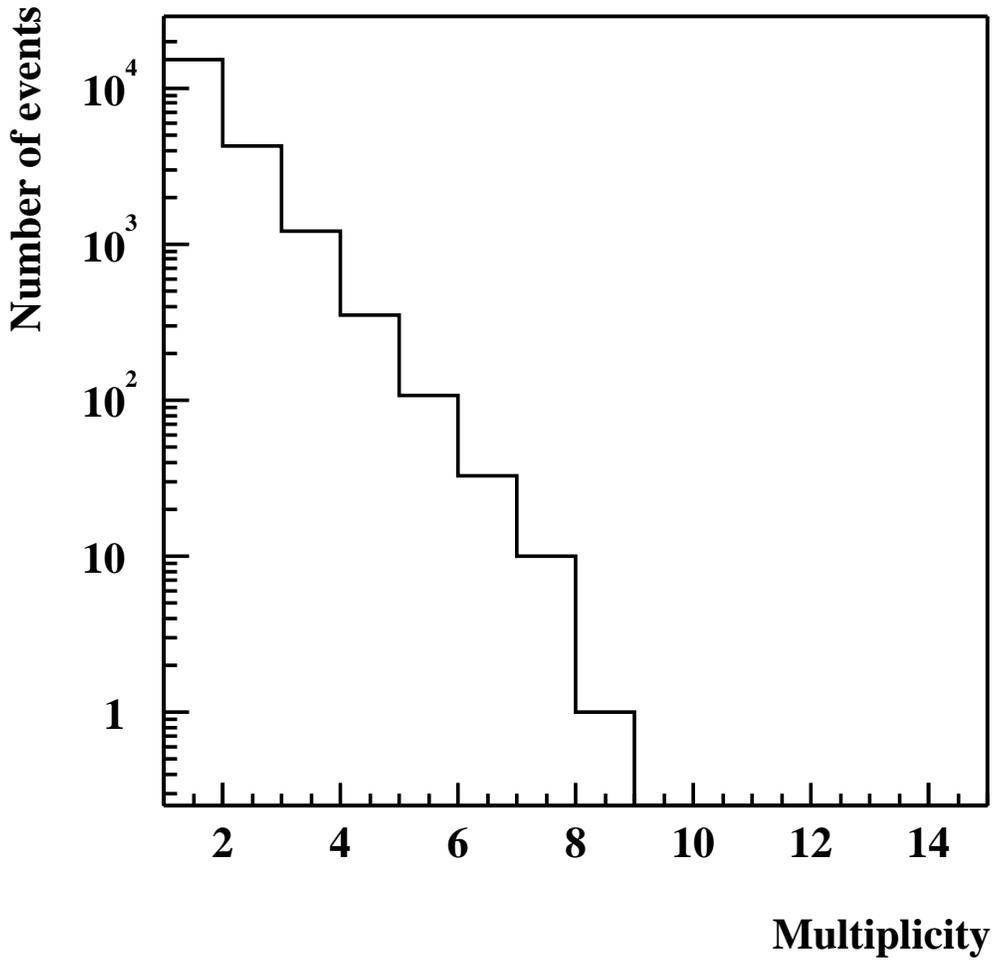,height=15cm}
\caption{Distribution of nuclear recoil multiplicities
in events caused by R8778 PMT neutron background in a 250 kg liquid xenon
detector. The first bin corresponds to single recoil events.
In this particular case an energy threshold of 10 keV for each 
nuclear recoil in an event (not for the sum of the recoil energies in 
an event) was applied.} 
\label{fig-mult}
\end{center}
\end{figure}

\begin{figure}[htb]
\begin{center}
\epsfig{figure=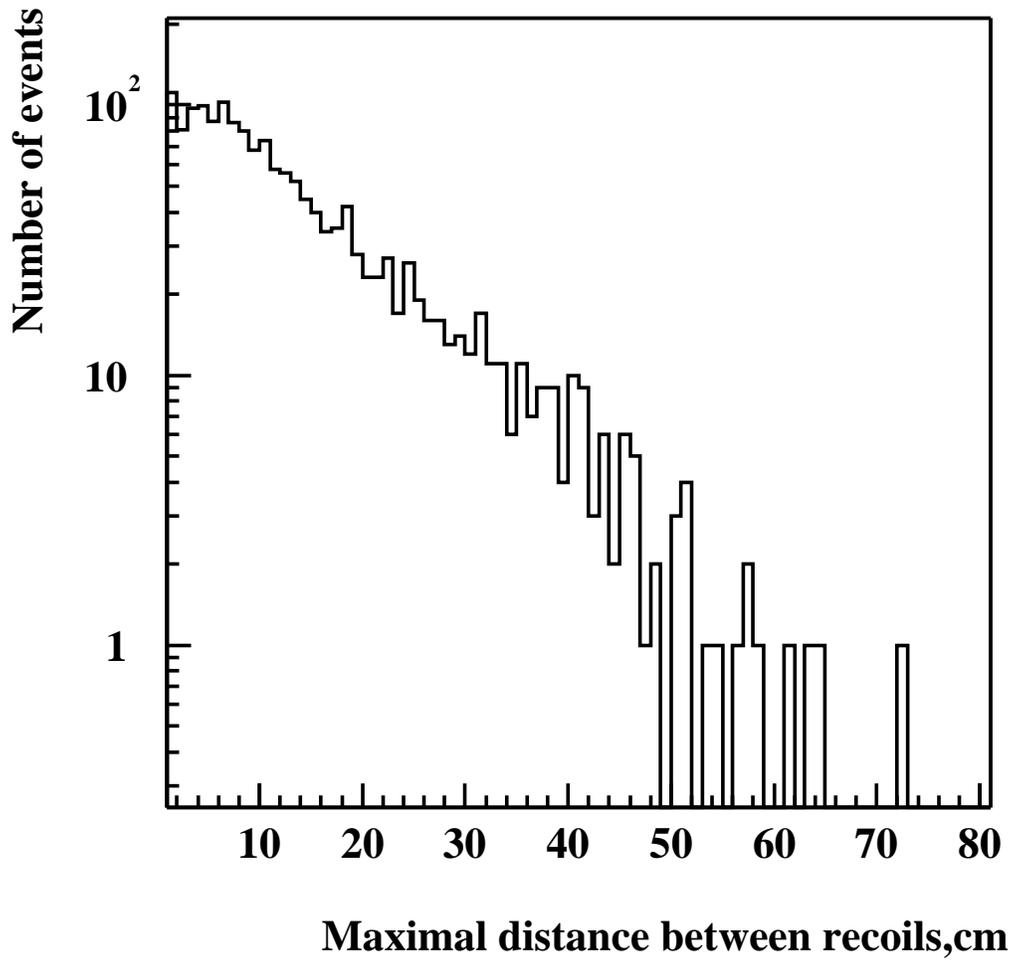,height=15cm}
\caption{Distribution of maximal distances between
nuclear recoils in events occuring in a 250 kg liquid xenon
detector due to neutrons from R8778 PMTs.} 
\label{fig-dist}
\end{center}
\end{figure}

Due to the very small interaction cross-section, a WIMP should produce
no more than one nuclear recoil per event. A neutron can produce
one or more recoils, but only events with a single recoil can mimic
WIMP interactions if a detector has position sensitivity. 
Figure \ref{fig-mult} shows the expected multiplicity
distribution of recoils (with an energy threshold of 10 keV for each 
recoil) for low background PMTs and
Figure \ref{fig-dist} shows the number of events with multiple
recoils as a function of maximal distance between recoils. Note that
multiple recoils here are produced by the same neutron, whereas
similar events from muon-induced neutrons can be due to several
neutrons generated in the same cascade.
The rejection of multiple recoil events not only reduces the background 
by more than 60\% (37\% of the initial value --
exact number depends on the position sensitivity of a detector) but also
provides a method of calculating the expected single recoil rate from 
neutrons. Single and
multiple recoil rates depend on the same factors, such as neutron
interaction cross-section, detector geometry and detector response.
Matching the calculated multiple recoil rate due to neutron 
interactions with measurements, we can then estimate the expected
single recoil rate due to neutrons and subtract it from the
measured single recoil rate. Note that a similar technique was applied
to subtract neutron-induced background in the CDMS experiment
\cite{cdms}.

Some improvements in the estimate of the neutron background 
may come from measurements of the energy
spectrum of recoils. This, however, needs further study, 
in particular detailed
comparison of simulated spectra from neutrons and from
WIMPs with various masses. Also it is not obvious that
appropriate statistics in real data will be available to
make this method work.

Finally, if the hydrocarbon material around a xenon detector is in fact
an active veto system, then some neutrons will give signals
in both the target and veto detectors, as discussed in
Ref. \cite{alex}. Rejecting those events
produced by neutrons, we can also use their rate 
and spectrum as a basis for calculating the single recoil
rate in xenon and again subtract an estimated neutron signal
from the measured recoil rate.

All these methods combined together should provide an
unambiguous estimate of the neutron-induced nuclear recoil rate
and grounds for statistical neutron background subtraction.
In the case of complete matching between the measured and
predicted background rates, the errors of each added in quadrature
result in a total error, which is proportional to the
square root of the measured rate. For example, having 300 events
during one year of experiment live time 
(as in the case of ETL 9266 PMTs) and expecting a similar number
from neutrons, we can calculate the mean value
of WIMP-induced events (with full background subtraction)
as around 0 with the standard deviation
of $\sqrt{2 \times 300} = 24.5$ and a limit at 90\% confidence
level (C.L.) of about 40. Hence, a factor of 8
reduction is expected with a full statistical subtraction 
of the neutron-induced background. For a measured number of 4 events
during 1 year of detector running time
the statistical tables \cite{pdg} can be used to obtain an upper limit
on a signal rate, which is 4.6 if the estimated background is
exactly equal to the measured rate.

\vspace{0.5cm}
{\large \bf 6. Bounds on WIMP sensitivity of large-scale 
xenon detectors from neutron background}
\vspace{0.3cm}

Events induced by neutrons in the
detector, which mimic WIMP-produced nuclear recoils, can be considered 
as a crucial factor that limits detector sensitivity.
Based on the above results it is worth estimating the limits on 
WIMP-nucleon cross-sections, which can be achieved with various rates
of neutron-induced events, as a means to aiding the design of future
detectors and improving their characteristics. 

Gamma background may also be a
limiting factor. Gamma background is higher by
several orders of magnitude (mainly due to internal contamination
of the target and, possibly, PMTs) than neutron background after 
shielding, but techniques are being developed, which give efficient
discrimination between electron and nuclear recoils. 
The rate of gamma events remaining after rejection depends on the
discrimination power, which is a characteristic of a particular
detector. Typically a factor of $10^7$ suppression is required to achieve
sensitivity down to $10^{-10}$ pb at the minimum of the sensitivity curve. 
However, the statistical
suppression factor, defined as the ratio of the gamma background rate to
the limit on the nuclear recoil rate, can be much higher than the discrimination
power deduced on an event-by-event basis, often called the figure of merit. 
Use of $^{85}$Kr free xenon
and other pure materials can soften this requirement by 2-3 orders
of magnitude. ZEPLIN I, a liquid xenon detector at Boulby, already 
achieved the statistical suppression factor of about $10^3$ \cite{z1-idm}.
Consideration of the gamma
background and discrimination power is, however, beyond the scope
of this work. We assume instead that gamma background is discriminated
against nuclear recoils so that it does not produce serious
problems for detector sensitivity.

Alphas can also contribute to background events at low energies.
Although they typically deposit a few MeV in the target, converted into
an MeV signal after correction for quenching, 
some of them can be emitted close to the
surface of the target and so lose energy in a keV range. 
For those originating within the xenon, it is possible for only a 
small fraction of their energy to be deposited inside the target
(accompanied by a nuclear recoil energy deposition), 
the remaining energy being lost in the vessel walls, 
where it is not detected.
In a similar way, alphas emitted in the vessel or PMTs close to the
surface can deposit a large fraction of their energy there,
enter the target and deposit a remaining small fraction in the xenon, which
will be detected. In both cases the events can be misinterpreted as
low energy nuclear recoil events, unless a topological study reveals 
that the events occured close to the surface of the target and/or the
discrimination between nuclear recoils and alphas allows complete rejection
of alpha events.
We will assume here that xenon purification, discrimination between
nuclear recoils and alphas, and topology study
will reduce alpha background down to a non-observable level.

Neutrons from radioactivity in rock and laboratory walls can be absorbed
down to a level of less than 0.2 events per year
by about 40 g/cm$^2$ thick slab of hydrocarbon placed behind lead 
shielding (between lead and detector).
In the case of the harder neutron spectrum, large thickness of 
CH$_2$ may be needed. This is the main result of Section 2.
So, the rate of events due to this source of background will be negligibly 
small. Radon can be kept away from the detector by using gas-tight sealing.

For muon-induced neutrons it is worth considering two cases: i) passive
hydrocarbon shielding around the detector, and ii) an active veto system
with liquid or plastic scintillator between lead and the detector (which can
act as gamma, neutron or muon veto). For case i) 7.5 events per year 
with measured energies between 2 and 10 keV are
expected in a 250 kg xenon detector. Note that some uncertainty remains
because of the simplified treatment of the nuclear recoil energy
in FLUKA. The rate can be as small as half that value if the event
topology study rejects most multiple recoil events.
This sample is a small fraction of events associated with
cosmic-ray muons. Providing the agreement between measurements and
simulations is reached for all these events, as well as for multiple
recoil events without electron recoils, a statistical subtraction
of this background can be done, improving the detector
sensitivity.
For case ii) less than 0.8 events per year (at 90\% C.L.) are expected 
independently
of the recoil spectrum (the accuracy is restricted by the statistics of 
the simulations).

The neutron background from detector components depends largely on the
readout components. With standard UV PMTs (with
quartz window and graded seal giving a high background -- 
not considered in detail here) 
the rate of background events
(several thousand events per year at 2-10 keV measured energy in a 
250 kg detector) is too high and allows only
a sensitivity of $10^{-7}$ pb to be reached at the minimum
of the sensitivity curve. 
Although statistical background subtraction can reduce this limit
by more than an order of magnitude, it is obvious
that these PMTs cannot be used in future large-scale
dark matter detectors. 

Low background ETL 9266 PMTs will produce about 300 events per year at 
2-10 keV measured energy in a 250 kg xenon detector. A factor of 8
reduction in limit can be achieved if the measured spectrum and 
multiplicity of these events agree with simulations, providing
a good reason for statistical subtraction of this background.
A liquid scintillator veto around the detector (instead of or 
in addition to the passive shielding) can be used in anticoincidence
with the main xenon detector allowing rejection of events with
nuclear recoils in both target and veto system. Again, accurate
simulations of the coincident events will help to estimate the
remaining background and statistically subtract it from the
measured rate. Ultra-low-background Hamamatsu R8778 PMTs would
yield background rates of 4-8 events per year depending
on whether neutron absorbing lightguides are used.
Active veto system around the target should
reject some fraction of events induced by neutrons from detector
components. The simulation of veto efficiency for these events
is the subject of a separate study.

Further improvements in PMTs discussed in the previous Section
can reduce the background rate to probably less than 1 event per year.
New readout techniques, such as GEM \cite{gem} or MICROMEGAS 
\cite{micromegas}, are of potential interest because they may
help to avoid using PMTs with their relatively high level of
radioactivity. Without PMTs the remaining neutron background comes mainly
from the copper vessel and gives a total rate
of less than 1 event per year.

Table \ref{table-sum} shows the summary of the neutron background
rates in a 250 kg xenon detector for 5 different detector
configurations as follows:\\
\noindent {\it Configuration 1}: 169 R8778 PMTs (4 ppb of U and Th), 
no lightguides, passive shielding around the detector 
(30 cm of lead and 40 g/cm$^2$ of hydrocarbon), single and multiple
recoils induced by neutrons.\\
\noindent {\it Configuration 2}: 169 R8778 PMTs (4 ppb of U and Th), 
no lightguides, passive shielding around the detector 
(30 cm of lead and 40 g/cm$^2$ of hydrocarbon), only single recoils
induced by neutrons are counted.\\
\noindent {\it Configuration 3}: 169 R8778 PMTs, no lightguides,
passive and active (to reject muon-induced neutrons)
shielding around the detector (30 cm of lead and 40 g/cm$^2$ of 
hydrocarbon), only single recoils are counted.\\
\noindent {\it Configuration 4}: 169 R8778 PMTs, acrylic lightguides,
passive and active shielding around the detector 
(30 cm of lead and 40 g/cm$^2$ of hydrocarbon), both single and multiple 
recoils are counted.\\
\noindent {\it Configuration 5}: large PMTs with 1 ppb of U and Th, 
acrylic lightguides,
passive and active shielding around the detector 
(we assumed 1 detected event instead of an average figure of 0.8
events per year if neutron contribution from rock and muons can be
neglected). 
Similar numbers are expected from
new readout techniques without PMTs, such as GEM or MICROMEGAS.

The rates at 2-10 keV electron equivalent energy range
(10-50 keV nuclear recoil energy) are shown in Table \ref{table-sum}.
Rock neutrons are absorbed to a level of less than 0.2 events/year by
passive and active shielding for all five configurations and are
neglected for all detector configurations, as well as muon-induced
neutrons for the configurations with an active veto system.
A copper vessel with a weight of 350 kg is presumed to
have 0.02 ppb of U and Th. Hydrocarbon material was placed
between PMTs in all configurations.
If a background rate can be estimated, as discussed in
Section 5, then it can be statistically subtracted, reducing
upper limits on nuclear recoils.

Figure \ref{fig-sens} summarizes the limitations on the sensitivity of a 
future 250 kg xenon detector running for a year for dark matter searches
assuming 100\% rejection of gamma and alpha induced events.
Although future tonne-scale detectors are aimed at discovering
WIMPs, we present the results in more conventional terms of 
limits on WIMP cross-section as a function of WIMP mass
to allow direct comparison between different existing and planned
detectors. 
An isothermal spherical 
halo with 0.3 GeV/cm$^3$
WIMP density and Maxwellian WIMP velocity distribution was assumed
and the procedure described by Lewin and Smith \cite{ls} was
followed. 
For any
specific detector configuration (see below) the parameter
space below the curve cannot be probed because of the neutron 
background. 
The dotted line shows
the limitations on the sensitivity for a detector with 169 2-inch low background 
R8778 PMTs surrounded by a hydrocarbon passive shielding (about 15.5 
background events per year 
at 2-10 keV measured electron equivalent energy from PMTs 
and muon-induced neutrons giving an upper limit of 21.9 events per year,
as for Configuration 1). 
The dashed line corresponds to a detector with a
scintillator veto system and all
the background statistically subtracted (3.0 events per year
for both measured and estimated background which corresponds to
an upper limit at 90\% C.L. of about 4.4 events per year,
as for Configuration 3). 
The solid line shows the limit achievable by a
detector with an active veto, lightguides and future new large PMTs 
with U/Th levels down to 1 ppb
(about 1 event per year 
giving less than 3.9 events per year at a 90\% C.L., as
for Configuration 5).
A new type of readout technique (GEMs or MICROMEGAS)
with very low levels of radioactivity (less than 1 event per year coming 
from the copper vessel) would provide a similar sensitivity. 
Finally the dashed-dotted curve shows the ultimate limit
for a detector with no background events observed during
one year of running, reachable with a charge readout technique
and ultra-pure components, for instance, less than 0.01 ppb of U/Th
in the copper vessel.
The above figures can be scaled up to a larger detector. 
Assuming 0.01 ppb of U and Th in a 350 kg copper vessel is the only
source of neutron background and a 250 kg liquid xenon detector
has an infinite running time, the nuclear recoil rate in the detector
is expected to be 0.2 events per year and the sensitivity to WIMP-nucleon
cross-section, which can be
achieved with such a rate without background subtraction, is about
$6 \times 10^{-12}$ pb at the minimum of the sensitivity curve.

A step function energy threshold of 2 keV was assumed in the
calculations of the sensitivities. This is, however, a specific
parameter for any particular experiment. Lowering the energy threshold
down to 1 keV may improve the sensitivity by about 50\% if 
no background events below 2 keV are detected.
In reality the trigger efficiency is not a step function
and should be both measured and calculated for a particular
detector. The energy resolution was taken as $1.24 \sqrt{E}$
as measured for the ZEPLIN I experiment \cite{zeplin1}. Note, however,
that for xenon experiments the sensitivity at the minimum of the curve
does not depend strongly on the energy resolution.

\vspace{0.5cm}
{\large \bf 7. Summary and conclusions}
\vspace{0.3cm}

Various sources of neutron background for a future large-scale
xenon dark matter detector have been investigated.
Shielding with 35-50 g/cm$^2$ of hydrocarbon (depending on
the thickness of lead shielding and the neutron spectrum) 
is needed to suppress
the neutron background from rock down to a level acceptable
for a high sensitivity dark matter detector (less than 1 event
per year in a 250 kg xenon detector).

Hydrocarbon is also very effective in suppressing the neutron
flux induced by cosmic-ray muons in rock and lead (if lead
shielding is placed between the rock and hydrocarbon).
Most events from muon-produced neutrons in the detector
will contain not only nuclear recoils but also muons,
photons and electrons associated with muon-induced cascades.
Remaining neutron events can be rejected by
anticoincidence with an active veto system placed around
the detector (the hydrocarbon can be the liquid scintillator
of an active veto system). Additional
rejection of neutron-induced nuclear recoil events is possible through
studying event topology, most neutron-induced
events being multiple recoil events.

Standard photomultiplier tubes with quartz window and graded seal
produce a high flux of 
neutrons from ($\alpha$,n) reactions and hence cannot be used
in detectors aimed to reach a sensitivity below $10^{-8}$ pb.
Ultra-low-background tubes can generate
about 4-8 nuclear recoil events per year in a 250 kg xenon
detector. Other detector components can
contribute a small fraction to the total background.
Further improvements in PMTs (larger size and lower
contamination levels) can bring this number down to
less than 1 event per year.

Restrictions on the detector sensitivity associated with the
neutron background have been studied. Neutrons from 
($\alpha$, n) reactions in rock and cosmic-ray muons do not
limit the sensitivity to WIMP-nucleon
interactions down to about $10^{-10}$ pb, especially if an active hydrocarbon 
veto (rather than passive shielding) is installed around the
detector. 
Some detector components, such as PMTs (even 
ultra-low-background PMTs) and associated equipment, 
produce a noticeable flux of neutrons.
This flux, however, does not limit detector sensitivity down to about 
$(1.3-1.5) \times 10^{-10}$ pb at the minimum of the sensitivity curve if
neutron absorbing lightguides are used, multiple recoils are rejected and 
statistical subtraction of background is possible 
based on comparison between measurements and simulations
of spectra, multiplicity and topology of background events.
Some events can be individually rejected by anticoincidence
with an active veto system. This needs further study. 

The best way to improve the sensitivity is to use larger diameter PMTs 
with lower contamination level 
or an alternative technique for charge readout
based on materials with an ultra-low level of radioactivity.

\vspace{0.5cm}
{\large \bf 8. Acknowledgments}
\vspace{0.3cm}

\indent The work has been undertaken within the framework of the
UK Dark Matter Collaboration (University of Edinburgh, 
Imperial College, London,
Rutherford Appleton Laboratory and University of Sheffield)
and contributes to the collaboration-wide simulation efforts.
This work is funded by PPARC. The authors are grateful to
Prof. P. F. Smith, Dr. N. J. T. Smith, 
Prof. T. J. Sumner, Prof. J. J. Quenby, Dr. J. D. Lewin, Dr. R. L\"uscher,
Dr. A. S. Howard, Dr. J. V. Dawson, Dr. H. Araujo, Dr. I. Liubarsky,
Dr. A. St. J. Murphy, Prof. D. P. Snowden-Ifft and Prof. C. J. Martoff
for valuable discussions.
Special thanks are addressed to Prof. P. F. Smith, who
performed comparison of our results with his simulations.
The authors acknowledge useful comments from an anonymous referee.

\pagebreak

\begin{table}[htb]
\caption{ Materials used in PMTs with their weights, 
contamination levels of U and Th, neutron production rates from U and Th 
impurities in PMTs,
and nuclear recoil rates per (kg $\times$ year) of target exposure (10-50
keV recoil energy)
from various components for ETL 9266 and Hamamatsu R8778.}
{\small
\begin{center}
\begin{tabular}{|c|c|c|c|c|c|c|c|}
\hline
Material & Mass & U & Th & Neutrons(U) & Neutrons(Th) & Rate(U) & Rate(Th) \\
& g & ppb & ppb & cm$^{-3}$ s$^{-1}$ &
cm$^{-3}$ s$^{-1}$ & kg$^{-1}$ y$^{-1}$ &kg$^{-1}$ y$^{-1}$ \\
\hline
ETL 9266 \\
\hline
Borosilicate & 70 & 30 & 30 
& $7.5 \times 10^{-9}$ & $2.3  \times 10^{-9}$ & 0.78 & 0.25 \\
\hline
Ceramics & 10 & 50 & 60 
& $1.45 \times 10^{-8}$ & $7.8 \times 10^{-9}$ & 0.1 & 0.08 \\
\hline
Metals & 20 & 0 & 30 & 0 & $4.5 \times 10^{-10}$ & 0 & 0.004 \\
\hline
R8778 \\
\hline
Quartz & 7 & 4 & 4 & 
$2.2 \times 10^{-10}$ & $4.8 \times 10^{-11}$ & 0.002 & 0.0004 \\
\hline
Metals & 153 & 4 & 4 & 
$4.8 \times 10^{-10}$ & $6.0 \times 10^{-11}$ & 0.024 & 0.004 \\
\hline
\end{tabular}
\end{center}
\label{table-ncomp}
}
\end{table}

\begin{table}[htb]
\caption{ Neutron background rates per year in a 250 kg liquid xenon detector
for 5 different configurations (see text for details).
Contributions from the main sources of neutron background are shown
separately together with the total rate.}
\begin{center}
\begin{tabular}{|c|c|c|c|c|c|}
\hline
Configuration & 1 & 2 & 3 & 4 & 5\\
\hline
Rock neutrons & $<$0.2 & $<$0.2 & $<$0.2 & $<$0.2 & $<$0.2 \\
\hline
Neutrons from muons & 7.5 & 3.3 & $<$0.8 & $<$0.8 & $<$0.8 \\
\hline
PMTs & 7.6 & 2.8 & 2.8 & 3.6 & 0.4 \\
\hline
Copper vessel & 0.4 & 0.2 & 0.2 & 0.4 & 0.4 \\
\hline
Total & 15.5 & 6.3 & 3.0 & 4.0 & 1.0\\
\hline
\end{tabular}
\end{center}
\label{table-sum}
\end{table}

\pagebreak

\pagebreak

\pagebreak

\pagebreak

\pagebreak

\pagebreak

\pagebreak

\pagebreak

\pagebreak

\pagebreak

\pagebreak

\pagebreak

\pagebreak

\pagebreak

\end{document}